\documentclass[preprint,prd,eqsecnum,aps,epsf]{revtex4}

\def\xslash#1{{\rlap{$#1$}/}}

\newcommand{\be}{\begin{eqnarray}}
\newcommand{\beq}{\begin{eqnarray}}
\newcommand{\befg}{\begin{figure}}
\newcommand{\CDP}{\hat{\cal D}\hspace{-0.27cm}\slash_v}
\newcommand{\CDPL}{\overleftarrow{\hat{\cal D}}\hspace{-0.29cm}\slash_v}
\newcommand{\CDPN}{{\cal D}\hspace{-0.26cm}\slash_v}
\newcommand{\CDPNL}{\overleftarrow{\cal D}\hspace{-0.29cm}\slash_v}

\newcommand{\DC}{D_{\bot}}
\newcommand{\DS}{D\hspace{-0.25cm}\slash}
\newcommand{\DSP}{\not\!{v} v\cdot D}
\newcommand{\DSPl}{\not\!{v} v\cdot \stackrel{\hspace{-0.1cm}\leftarrow} D}
\newcommand{\DSPR}{\not\!{v} v\cdot \partial}
\newcommand{\DSPL}{\not\!{v} v\cdot \stackrel{\hspace{-0.1cm}\leftarrow}\partial}
\newcommand{\DSp}{\not\!{v} v\cdot \partial}
\newcommand{\dsp}{D\hspace{-0.27cm}\slash_{\|}}
\newcommand{\DSC}{D\hspace{-0.26cm}\slash_{\bot}}
\newcommand{\KSC}{k\hspace{-0.26cm}\slash_{\bot}}
\newcommand{\KC}{k_{\bot}}
\newcommand{\DP}{D_{\|}}
\newcommand{\DSCX}{D\hspace{-0.26cm}\slash_{\bot}}
\newcommand{\DSCXL}{\overleftarrow{D}\hspace{-0.29cm}\slash_{\bot}}

\newcommand{\edfg}{\end{figure}}
\newcommand{\eeq}{\end{eqnarray}}
\newcommand{\ee}{\end{eqnarray}}

\newcommand{\MQ}{m_Q}
\newcommand{\MQP}{m_{Q^{\prime}}}

\newcommand{\oDSP}{\not\!{v} v\cdot \overleftarrow{D}}
\newcommand{\oDSp}{\not\!{v} v\cdot \overleftarrow{\partial}}
\newcommand{\odsp}{\overleftarrow{D}\hspace{-0.29cm}\slash_{\|}}
\newcommand{\oDslashbot}{\overleftarrow{/\!\!\!\!D}\hspace{-5pt}_{\bot}}

\newcommand{\PP}{{1 + v\hspace{-0.2cm}\slash \over 2}}
\newcommand{\PM}{{1 - v\hspace{-0.2cm}\slash \over 2}}

\newcommand{\qpv}{Q_v^{(+)}}
\newcommand{\qpvb}{\bar{Q}_v^{(+)}}
\newcommand{\QVBP}{\bar{Q}^{(+)}_{v^{\prime}} }
\newcommand{\QV}{Q^{(+)}_v}
\newcommand{\QVB}{\bar{Q}^{(+)}_v}
\newcommand{\Qv}{Q_v}
\newcommand{\QvB}{\bar{Q}_v}
\newcommand{\QvBP}{\bar{Q}'_{v^{\prime}} }
\newcommand{\QVH}{\hat{Q}_v}
\newcommand{\QVHB}{\bar{\hat{Q}}_v}
\newcommand{\QVHF}{\hat{Q}^{(-)}_v}
\newcommand{\QVHP}{\hat{Q}^{(+)}_v}
\newcommand{\QVHMP}{\hat{Q}^{(\mp)}_v}
\newcommand{\QVHPB}{\bar{\hat{Q}}_v{\vspace{-0.3cm}\hspace{-0.2cm}{^{(+)}} }}
\newcommand{\QVHFB}{\bar{\hat{Q}}_v{\vspace{-0.3cm}\hspace{-0.2cm}{^{(-)}} }}
\newcommand{\QVHPM}{\hat{Q}^{(\pm)}_v}

\newcommand{\QVHPMB}{\bar{\hat{Q}}_v{\vspace{-0.3cm}\hspace{-0.2cm}{^{(\pm)}} }}
\newcommand{\QVHZB}{\bar{\hat{Q}}_v{\vspace{-0.3cm}\hspace{-0.2cm}{^{(+)}} } }
\newcommand{\QVHZ}{\hat{Q}^{(+)}_v}

\newcommand{\uvslash}{/\!\!\!\!\hspace{1pt}v}
\newcommand{\VS}{\not\!{v}}

\newcommand{\DL}{\stackrel{\leftarrow}{D}}
\newcommand{\DCLS}{\stackrel{\hspace{-0.02cm}\leftarrow}{D_{\bot}^2}}

\begin{document}
\title{Large Component QCD and Theoretical Framework of \\ Heavy Quark Effective Field Theory}
\author{ Yue-Liang Wu }
\address{ Institute of Theoretical Physics, Chinese Academy of Sciences, Beijing 100080, China }
\begin{abstract}
Based on a large component QCD derived directly from full QCD by
integrating over the small components of quark fields with $|{\bf
p}| < E + m_Q$, an alternative quantization procedure is adopted
to establish a basic theoretical framework of heavy quark
effective field theory (HQEFT) in the sense of effective quantum
field theory. The procedure concerns quantum generators of
Poincare group, Hilbert and Fock space, anticommutations and
velocity super-selection rule, propagator and Feynman rules,
finite mass corrections, trivialization of gluon couplings and
renormalization of Wilson loop. The Lorentz invariance and
discrete symmetries in HQEFT are explicitly illustrated. Some new
symmetries in the infinite mass limit are discussed. Weak
transition matrix elements and masses of hadrons in HQEFT are well
defined to display a manifest spin-flavor symmetry and $1/m_Q$
corrections. A simple trace formulation approach is explicitly
demonstrated by using LSZ reduction formula in HQEFT, and shown to
be very useful for parameterizing the transition form factors via
$1/m_Q$ expansion. As the heavy quark and antiquark fields in
HQEFT are treated on the same footing in a fully symmetric way,
the quark-antiquark coupling terms naturally appear and play
important roles for simplifying the structure of transition matrix
elements, and for understanding the introduction of `dressed heavy
quark' - hadron duality. In the case that the `longitudinal' and
`transverse' residual momenta of heavy quark are at the same order
of power counting, HQEFT provides a consistent approach for
systematically analyzing heavy quark expansion in terms of
$1/m_Q$. Some interesting features in applications of HQEFT to
heavy hadron systems are briefly outlined.
\end{abstract}

 \pacs{ 12.38.-t, 12.39.-x, 11.10.-z}
\maketitle

\newpage
{\bf Contents}
\\
\\
1.\ Introduction \\
2.\ {\bf  Large Component QCD (LCQCD) } \\
\  $\quad$ 2.1\ Effective Lagrangian of LCQCD from full QCD \\
\  $\quad$ 2.2\ Lorentz Invariance of LCQCD\\
\  $\quad$ 2.3\ Quark-antiquak couplings and their decouple
        conditions \\
\  $\quad$ 2.4\ Heavy Quark Effective Field Theory (HQEFT) from LCQCD \\
3.\ {\bf  Quantization of HQEFT } \\
\  $\quad$  3.1\ Quantum Generators of Poincare Group \\
\  $\quad$  3.2\ Anticommutations and Georgi's Velocity Superselection Rule \\
\  $\quad$  3.3\ Hilbert and Fock Space of HQEFT \\
\  $\quad$  3.4\ Propagator in HQEFT \\
\  $\quad$  3.5\ Discrete Symmetries in HQEFT  \\
4.\ {\bf  Basic Framework of HQEFT } \\
\  $\quad$  4.1\ Feynman Rules in HQEFT  \\
\  $\quad$  4.2\ Effective Lagrangian of HQEFT with Finite Mass Corrections \\
\  $\quad$  4.3\ Trivialization of Gluon Couplings and Decouple Theorem \\
\  $\quad$  4.4\ Renormalization in HQEFT and Wilson Loop \\
\  $\quad$  4.5\ Current Operators in HQEFT \\
\  $\quad$  4.6\  Spin and Angular Momentum in HQEFT \\
\  $\quad$  4.7\ New Symmetries in HQEFT at Infinite Mass Limit \\
5.\ {\bf Hadronic Matrix Elements and $1/m_Q$ Expansions in HQEFT } \\
\  $\quad$  5.1\ Weak Transition Matrix Elements in HQEFT \\
\  $\quad$  5.2\ Mass Formula of Hadrons and Transition Matrix Elements at Zero Recoil \\
\  $\quad$  5.3\ Trace Formula of Transition Matrix Elements and
Universal Isgur-Wise Function \\
\  $\quad$  5.4\ Transition Form Factors and $1/m_Q$ Expansions in HQEFT \\
\  $\quad$  5.5\ Interesting Features in Applications of HQEFT \\
6.\ {\bf Conclusions and Remarks }

\newpage

\section{Introduction}

Great experimental and theoretical progresses have been achieved
in heavy quark physics\cite{FF,AA}. As more and more accurate
experimental data become available in the future based on
B-factories and other colliders, it requires more precise
theoretical calculations for hadronic matrix elements since
physical observables can only be measured from hadronic physics
due to quark confinement. Thus one of the important issues
concerns the treatment of nonperturbative QCD as heavy hadrons are
bound states of quarks with a typical binding energy $
\bar{\Lambda} \sim 2\Lambda_{QCD} \simeq 600 $ MeV. The high
energy behavior of QCD is well understood from the asymptotic
freedom of QCD. To understand low energy dynamics of QCD, one
needs to develop an appropriate method for treating
non-perturbative QCD effects at low energies. As the relative
momentum of heavy quark within the hadrons is at the order of
binding energy scale which is smaller than the heavy quark mass.
For heavy bottom and charm quarks, their masses $m_Q = m_b,\ m_c$
are much larger than the QCD energy scale $m_Q \gg \bar{\Lambda}
\sim 2\Lambda_{QCD}$. In order to study low energy behavior of
heavy quark, it is useful to express the four momentum of heavy
quark as $p = m_Q v + k$ with $v^2 =1$. Here $k$ denotes the
residual momentum or relative momentum when $\nu_{\mu}$ is taken
to be the velocity of heavy hadron at the rest frame
$v=(1,0,0,0)$. As a heavy quark within the hadron has a typical
residual momentum $|{\bf k}| \sim \bar{\Lambda} \ll m_Q$, which
motivates us to develop an effective field theory to treat the
non-perturbative QCD effects for heavy quarks. A large component
QCD (LCQCD) treating large component effective heavy quark and
antiquark fields on the same footing can be derived directly from
full QCD at $|{\bf p}| \ll E + m_Q$ or $|{\bf k}| \ll 2m_Q + k^0$.
For the case that the heavy quark is nearly on-mass shell, i.e.,
$k^0 \sim |{\bf k} |^2/2m_Q $, LCQCD is shown to behave like a
non-relativistic QCD (NRQCD). When the heavy quark goes to
off-mass shell with $k^0  \gg |{\bf k} |^2/2m_Q$ (i.e., $|{\bf k}|
\ll \sqrt{2 k^0 m_Q} \sim \sqrt{2\bar{\Lambda}m_Q}$  for $k^0 \sim
\bar{\Lambda}$), the typical case occurs in the heavy-light hadron
system in which the ``longitudinal" and ``transverse" residual
momenta of heavy quark are at the same order of power counting
(i.e., $ k^0 \sim |{\bf k} | \sim \bar{\Lambda} \ll m_Q$), then
LCQCD can be treated as a heavy quark effective field theory
(HQEFT)\cite{ylw} via the heavy quark expansion in terms of
$1/m_Q$, which has been applied to deal with heavy-light hadron
systems\cite{W1,W2,W3,W4,W5,W6}. The leading term of effective
lagrangian in HQEFT characterizes the behavior of heavy quarks in
the infinite mass limit\cite{limit1,limit2} and coincides with the
standard one\cite{HG} which explicitly displays the heavy quark
spin-flavor symmetry\cite{sym1,sym2,IW}.

The effective Lagrangian in HQEFT\cite{ylw} differs from the
widely used one in the literatures \cite{MN} which was shown to be
constructed based on the assumption that particle and antiparticle
numbers are separately conserved \cite{MRR}. Thus the differences
between two effective theories arise mainly from the
quark-antiquark mixed interaction terms. Therefore, it is of
interest to show explicitly their differences and relations.
Namely, in which case the effective Lagrangian in HQEFT can
recover the widely used heavy quark effective Lagrangian.
Theoretically, it is known from quantum field theories that only
in the infinity mass limit the quark and antiquark numbers become
separately conserved as the quark and antiquark interactions
approach to decouple. In other words, with finite quark mass in
the real world, it requires one to keep the quark-antiquark
interaction terms and to treat the quark and antiquark fields on
the same footing in the sense of effective quantum field theory at
low energies. On the other hand, the effects of quark-antiquark
interaction terms at low energies should also distinguish from the
Wilson coefficient functions which mainly reflect the perturbative
QCD corrections at relatively high energy scales.

As LCQCD is directly derived from QCD, all contributions of the
field components, large and small, `particle' (positive energy
part) and `antiparticle' (negative energy part), have carefully
been dealt with in the effective Lagrangian. Especially, the
effective heavy quark and antiquark fields are treated on the same
footing in a fully symmetric way, the resulting effective
Lagrangian shall provide a complete description for both quark and
antiquark fields. In fact, HQEFT based on LCQCD has lead to a
number of successful phenomenological applications, such as: a
reliable extraction for the CKM matrix elements $|V_{cb}|$ and
$|V_{ub}|$\cite{W1,W2,W3}, a consistent explanation for the
lifetime differences among bottom hadrons\cite{W4},  a good
approximated scaling law for the decay constants of heavy
mesons\cite{W5}, an interesting observation for useful relations
of form factors in heavy to light hadron transitions\cite{W6}.  It
then strongly motivates us to present a complete theoretical
framework of HQEFT in the sense of effective quantum field theory
at low energies when the ``longitudinal" and ``transverse"
residual momenta of heavy quark are at the same order of power
counting in the $1/m_Q$ expansion, and to demonstrate a systematic
approach for evaluating the symmetry breaking effects due to
$1/m_Q$ corrections. This comes to our main purpose in the present
paper.

The paper is organized as follows: in section 2, we present a
complete derivation for the effective Lagrangian of LCQCD directly
from full QCD by integrating over the small components of quark
and antiquark fields, it is explicitly shown how the large
component quark-antiquark coupling terms are naturally resulted in
the effective Lagrangian of LCQCD. The Lorentz invariance of LCQCD
is explicitly illustrated, which leads to a super-equivalence of
LCQCD with different velocities.  We then discuss the special
cases in which the quark components or antiquark components
decouple from effective Lagrangian, so that the effective
Lagrangian is reduced to the widely used one in the literatures.
Our special attention is paid to consider two interesting cases
which are corresponding to two energy regions of heavy quark. We
will show how the two energy regions make LCQCD to be dealt with
as two effective theories: one like a non-relativistic QCD
(NRQCD), and another as a heavy quark effective field theory
(HQEFT). A detailed discussion is given for HQEFT with large
component quark-antiquark coupling terms. Of particular, we
explicitly demonstrate the important effects of quark-antiquark
coupling terms in HQEFT when the residual energy and momentum are
at the same order of power counting, i.e., $ k^0 \sim |{\bf k} |
\sim \bar{\Lambda} \ll m_Q$. The quantization of HQEFT is carried
out in section 3, as the large component effective heavy quark and
antiquark fields $Q_v (x)$ and $\bar{Q}_v (x)$ concern the
additional arbitrary unit time-like vector $v_{\mu}$, the
canonical quantization is not quite clear for it, we then adopt an
alternative quantization procedure, which concerns quantum
generators of Poincare group, Hilbert and Fock space,
anticommutations and velocity superselection rule, propagator and
discrete symmetries in HQEFT. In section 4, a basic theoretical
framework of HQEFT is established based on the quantization of
HQEFT and Feynman rules as well as the finite mass corrections in
the heavy quark expansion of $1/m_Q$. The trivialization of gluon
couplings and renormalization of Wilson loop are explicitly
demonstrated. Especially, the current operators via $1/m_Q$
expansion in HQEFT are discussed in detail. It is emphasized that
in the computation of the coefficient functions at sub-leading
order in $1/m_Q$, the large component quark-antiquark mixing terms
in the current operators must be kept as they will receive
contributions from the terms of the quark-antiquark coupling terms
in the effective Lagrangian via virtual quark or antiquark field
exchanges. It is also shown that there actually exist more
symmetries in HQEFT at infinite mass limit, in addition to the
well-known spin-flavor and angular momentum symmetries. It section
5, the weak transition matrix elements in HQEFT are well defined
to display the manifest spin-flavor symmetry and to exhibit a
systematic heavy quark expansion in terms of $1/m_Q$. All relevant
operators up to order of $1/m_Q^2$ are presented explicitly. An
alternative definition for the mass of heavy hadron is provided in
HQEFT via $1/m_Q$ expansion. Some important transition matrix
elements at zero recoil in HQEFT are investigated, it is shown
that $1/m_Q$ order corrections are automatically absent at zero
recoil point in HQEFT even without analyzing the concrete Lorentz
structure and the form factors of the transition matrix elements.
Of particualr, a simple trace formulation for evaluating the
transition matrix elements is explicitly demonstrated by using the
LSZ reduction formula for heavy hadron states in HQEFT. As a
consequence, the universal Isgur-Wise function is shown to relate
explicitly the overlapping integral between wave functions of
initial and final state mesons. We then show that the simple trace
formulation approach becomes very powerful for parameterizing the
transition form factors via $1/m_Q$ expansion in HQEFT.
Especially, as the effective quark and antiquark fields are
treated symmetrically on the same footing, it simplifies the
structure of transition matrix elements in the heavy quark
expansion in terms of $1/m_Q$. This is because the operators
appearing in the effective Lagrangian and effective current
contain only terms with even powers of ${\DSC}$ when the
contributions of quark-antiquark interaction terms in HQEFT are
considered. As HQEFT can be applied to treat off-mass shell heavy
quarks, the concept of ``dressed heavy quark" with well defined
dressed quark mass $\hat{m}_Q = m_Q + \bar{\Lambda} = \lim_{m_Q\to
\infty} m_H$ is introduced to characterize the confining binding
effects of non-perturbative QCD at low energies, here $m_H =
\hat{m}_Q \left(1 + O(1/m_Q^2) \right)$  is the heavy hadron mass.
The `dressed heavy quark'-hadron duality is found to be more
reasonable than the naive quark-hadron duality. Consequently,
HQEFT allows us to make a systematic and consistent evaluation for
the symmetry breaking corrections caused by the finite mass of
heavy quark. Some interesting features and phenomena for the
application of HQEFT are briefly outlined. Our conclusions and
remarks are presented in the final section.

\section{Large Component QCD }

\subsection{Effective Lagrangian of LCQCD from full QCD}

For completeness, we follow ref.\cite{ylw} to derive in a more
systematic way the effective lagrangian of LCQCD from full QCD.
The Lagrangian in full QCD is
\begin{equation}
\label{QCDL} {\cal L}_{QCD}= \bar{Q} (i\DS-m_Q) Q
 + {\cal L}_{light}
\end{equation}
where $Q$ denotes the heavy quark field and ${\cal L}_{light}$
represents the remaining part. It is well known that the Dirac
fermion fields contain particle and antiparticle components which
correspond to positive and negative energy parts. Formally, one
can always write the field $Q$ into two parts
\begin{equation}
    Q=Q^{(+)}+Q^{(-)} ,
\end{equation}
with $Q^{(+)}$ and $Q^{(-)}$ relating mainly to the positive energy
part and negative energy part of classical solutions respectively.
In the case of the free quark field, they are corresponding to two
solutions of Dirac equation
\begin{equation}
     \label{eqmotion}
(i\partial\hspace{-0.25cm}\slash-m_Q)Q^{(\pm)}=0.
\end{equation}
Here $Q^{(+)}$ and $Q^{(-)}$ are known as ``quark" and ``antiquark"
fields respectively with the following explicit solutions
\begin{eqnarray}
Q^{(+)}(x) &=& \int \frac{d^3p}{(2\pi)^3}{m\over p^0}
    \sum\limits_s b_s(p)u_s(p)e^{-ip\cdot x},  \\
Q^{(-)}(x) &=& \int \frac{d^3p}{(2\pi)^3}{m\over p^0}
    \sum\limits_s d^{\dag}_s(p)v_s(p)e^{ip\cdot x},
\end{eqnarray}
where $s$ is spin index, $b_s$ and $d_s$ are the annihilation and
creation operators respectively. $u_s$ and $v_s$ are
four-component spinors. In the Dirac representation, they are
given by the following explicit forms
\begin{eqnarray}
u_s(p) &=& \sqrt{E+m_Q \over 2m_Q} \left( \begin{array}{c} 1 \\
{{\bf \sigma \cdot p}\over E+m_Q} \end{array} \right)
\varphi_s ,  \\
v_s(p) &=& \sqrt{E+m_Q \over 2m_Q} \left( \begin{array}{c} {{\bf
\sigma \cdot p}\over E+m_Q} \\ 1 \end{array} \right) \chi_s
\end{eqnarray}
with $\varphi_s$ being the two component Pauli spinor field that
annihilates a heavy quark, and $\chi_s$ being the Pauli spinor
field that creates a heavy antiquark.

Introducing the projection operators $P_{\pm} =(1 \pm \uvslash)/2$
satisfying $P_{\pm}^2 = 1$ with $v^\mu$ being an arbitrary unit
vector $v^2 = 1$, we can decompose the quark fields $Q^{(\pm)}$
into the following forms
\begin{eqnarray}
\label{Qdec1} Q^{(+)} &\equiv& \Big({1 + \uvslash \over 2} + {1 -
\uvslash \over 2} \Big) Q^{(+)}
= \hat{Q}^{(+)}_v + R^{(+)}_v,  \nonumber \\
\label{Qdec2} Q^{(-)} &\equiv& \Big({1 - \uvslash \over 2} + {1 +
\uvslash \over 2}\Big) Q^{(-)} = \hat{Q}^{(-)}_v + R^{(-)}_v
\end{eqnarray}
with
\begin{eqnarray}
\hat{Q}^{(\pm)}_v\equiv
\frac{1{\pm}v\hspace{-0.2cm}\slash}{2}Q^{(\pm)}, \qquad
R^{(\pm)}_v\equiv \frac{1{\mp}v\hspace{-0.2cm}\slash}{2}Q^{(\pm)}
\end{eqnarray}

 Thus the initial quark field $Q$ can be rewritten as
 \begin{equation}
 Q \equiv \hat{Q}_v + R_v
 \end{equation}
 with
 \begin{equation}
 \hat{Q}_v = \hat{Q}_v^{(+)} + \hat{Q}_v^{(-)},
 \qquad R_v = R_v^{(+)} + R_v^{(-)} .
 \end{equation}
For free quark case, taking $v=(1, 0, 0, 0)$, the projected quark
fields get the following forms in the momentum space
\begin{eqnarray}
\label{component1} \hat{Q}_v^{(+)} \to \PP u_s(p) &=& \sqrt{E+m_Q
\over 2m_Q} \left( \begin{array}{c} 1 \\ 0 \end{array} \right)
\varphi_s ,
\\
\label{component2} R_v^{(+)} \to \PM u_s(p) &=& \sqrt{E+m_Q \over
2m_Q} \left( \begin{array}{c} 0 \\ {{\bf \sigma \cdot p}\over
E+m_Q}
\end{array} \right) \varphi_s ,
\\
\label{component3} R_v^{(-)} \to \PP v_s(p) &=& \sqrt{E+m_Q \over
2m_Q} \left( \begin{array}{c}{{\bf \sigma \cdot p}\over E+m_Q} \\
0
\end{array} \right) \chi_s ,
\\
\label{component4} \hat{Q}^{(-)}_v \to \PM v_s(p) &=& \sqrt{E+m_Q
\over 2m_Q} \left( \begin{array}{c} 0 \\ 1 \end{array} \right)
\chi_s .
\end{eqnarray}
which shows that in the case $ |{\bf p}| \ll E+m_Q $ the effective
heavy fields $\hat{Q}_v^{(+)}$ and $\hat{Q}_v^{(-)}$ are the
corresponding ``large components" of ``quark" and ``antiquark"
fields, while $R_v^{(+)}$ and $R_v^{(-)}$ are the corresponding
``small components" .

It is also useful to decompose the derivative operator into two
parts
\begin{eqnarray}
 \DS = \dsp +  \DSC
\end{eqnarray}
with
\begin{equation}
 \dsp \equiv \DSP, \quad \DSC \equiv \DS\  -\VS (v\cdot D) ,
\end{equation}
They satisfy the commutation relations
\begin{eqnarray}
 \label{commutation}
[\VS, \dsp]=0, \qquad \{\VS,  \DSC \}=0,
\end{eqnarray}
 Using this property, we may relate the field components $R^{(\pm)}_v$
($\bar{R}^{(\pm)}_v$) with $\QVHPM$ ($\QVHPMB$) via the following
equations:
\begin{eqnarray}
\label{RinQVH}
(i\dsp - m_Q) R_v^{(\pm)}  + i\DSC \hat{Q}_v^{(\pm)} = 0 , \\
\bar{R}^{(\pm)}_v (-i\odsp - m_Q)  - \QVHPMB i\oDslashbot = 0 .
\end{eqnarray}
which are actually the half part of Dirac equation of motion. With
this relations, $R_v$ becomes $1/m_Q$ suppressed relative to
$\QVH$. Thus $\QVHPM$ and $R^{(\pm)}_v$ can be regarded as the
``large components" and ``small components" of the heavy quark
fields $Q^{(\pm)}$ respectively. In this representation, it is
independent of the special choice of vector $v^{\mu}$ as long as
it satisfies the condition $v^2=1$. Note that one should not apply
the whole Dirac equation of motion to both quark and antiquark
components, otherwise the resulting effective field Lagrangian
vanishes. Namely we shall consider the off-mass shell cases with
\begin{eqnarray}
(i\dsp - m_Q) \hat{Q}_v^{(\pm)}   + i\DSC R_v^{(\pm)} \neq 0 , \\
 \QVHPMB (-i\odsp - m_Q)  -  \bar{R}^{(\pm)}_v i\oDslashbot \neq 0 .
\end{eqnarray}

 With the relation in (2.19) and (2.20), the fields $Q$ and $\bar{Q}$ can be represented
by $\hat{Q}_v$ and $\bar{\hat{Q}}_v$ via the following relations
\begin{eqnarray}
     \label{QinQVH1}
& &  Q  = \Big[1+\Big(1-\frac{i\DSP+\MQ}{2\MQ}\Big)^{-1}
           \frac{i\DSCX}{2\MQ}\Big] \hat{Q}_{v}\equiv \Big[ 1 + W_v (\frac{1}{m_Q}) \big]
           \hat{Q}_{v}
            \equiv \hat{\omega} \hat{Q}_v \\
 \label{QinQVH2}
& &    \bar{Q} = \bar{\hat{Q}}_{v}\Big[1- \frac{i \DSCXL }{2\MQ}
\Big(1+\frac{i\oDSP-\MQ}{2\MQ}\Big)^{-1}\Big] \equiv \Big[ 1 +
\overleftarrow{W}_v (\frac{1}{m_Q}) \Big]\bar{\hat{Q}}_v \equiv
\bar{\hat{Q}}_v\overleftarrow {\hat{\omega}}
\end{eqnarray} Here
for any operator $O$, the operator $\overleftarrow{O}$ is defined
via $\int\kappa\overleftarrow{O}\varphi \equiv -\int\kappa O
\varphi$.

 To construct the effective field theory of large component
$\hat{Q}_v$, one can simply integrate out the small component
$R_v$. At tree level, it is easily shown that this is equivalent
to substitute Eqs.(\ref{QinQVH1}) and (\ref{QinQVH2}) into the QCD
Lagrangian. As a consequence, we arrive at an effective lagrangian
for a large component QCD (LCQCD)
    \begin{equation}
    \label{eq:2.22}
       {\cal L}_{QCD} ={\cal L}_{light}+ \hat{{\cal L}}_{Q,v}
    \end{equation}
with
    \begin{eqnarray}
    \label{eq:2.26}
     & &  \hat{{\cal L}}_{Q,v} \equiv \bar{Q}(i\DS-\MQ)Q |_{Q\to \hat{\omega}\hat{Q}_v}  \nonumber \\
   & & = \QVHB[i\CDP -\MQ]\QVH
 +\frac{1}{2\MQ}\QVHB (i \CDPL -\MQ)
\Big(1-\frac{i\DSP+\MQ}{2\MQ}\Big)^{-1} (i\DSCX )
   \QVH  \\
   & & \equiv  \QVHB[i\CDP -\MQ]\QVH
  +\frac{1}{2\MQ}\QVHB (-i\DSCXL)\Big(1-\frac{-i\oDSP+\MQ}{2\MQ}
   \Big)^{-1}(i\CDP -\MQ) \QVH \nonumber
    \end{eqnarray}
where the operator $i\CDP$ is defined as
\begin{eqnarray}
i\CDP &=& i\DSP + \frac{1}{2\MQ}i\DSC
\Big(1-\frac{i\DSP+\MQ}{2\MQ} \Big)^{-1} i\DSC .
\end{eqnarray}

 It is easily seen from the definitions (eqs.2.8-2.11) and the commutation relations
(\ref{commutation}) that there is no quark-antiquark coupling in
the first term on the rhs. of Eq.(\ref{eq:2.26}). The second term
on the rhs. of Eq.(\ref{eq:2.26}) arises from the quark-antiquark
coupling interactions. To be manifest, one can rewrite the above
Lagrangian ${\cal L}_{Q,v}$ in the following form
   \begin{equation}
     \label{lagr}
        \hat{{\cal L}}_{Q,v} = \hat{{\cal L}}^{(++)}_{Q,v}+\hat{{\cal L}}^{(--)}_{Q,v}
          +\hat{{\cal L}}^{(+-)}_{Q,v}+\hat{{\cal L}}^{(-+)}_{Q,v}
     \end{equation}
with
    \begin{eqnarray}
\hat{{\cal L}}^{(\pm \pm)}_{Q,v} &=&
        \QVHPMB [i\CDP -\MQ] \QVHPM , \\
\hat{{\cal L}}^{(\pm \mp)}_{Q,v} & = & \frac{1}{2\MQ} \QVHPMB (i
\CDPL -\MQ) \Big(1-\frac{i\DSP+\MQ}{2\MQ}\Big)^{-1} (i\DSCX
)\QVHMP \\
 & \equiv & \frac{1}{2\MQ}
\QVHPMB (-i\DSCXL)\Big(1-\frac{-i\oDSP+\MQ}{2\MQ}
   \Big)^{-1}(i\CDP -\MQ) \QVHMP,
\end{eqnarray}
The operator $\CDPL$ in the above equations can be obtained by
replacing $D^{\mu}$ with $-\overleftarrow{D^{\mu}}$ in $\CDP$.

To make $1/m_Q$ expansion, it is useful to remove the large mass
term in the effective Lagrangian. We then introduce new field
variables $Q_v$ and $\bar{Q}_v$ with the definition
    \begin{equation}
       Q_v=e^{iv\hspace{-0.15cm}\slash m_Q v\cdot x}\QVH ,\hspace{1.5cm}
       \bar{Q}_v=\QVHB e^{-i v\hspace{-0.15cm}\slash m_Q v \cdot x} .
    \end{equation}
Noticing the feature that $\VS$ commutes with $\dsp$ but
anticommutes with $\DSC$,  we can rewrite above Lagrangian to be
\begin{eqnarray}
 {\cal L}_{Q,v} & = & {\cal L}^{(1)}_{Q,v} + {\cal L}^{(2)}_{Q,v}
 \nonumber \\
 {\cal L}^{(1)}_{Q,v} & \equiv & {\cal L}^{(++)}_{Q,v} + {\cal
L}^{(--)}_{Q,v} =
        \bar{Q}_v i\CDPN Q_v , \\
{\cal L}^{(2)}_{Q,v} & \equiv & {\cal L}^{(+-)}_{Q,v} + {\cal
L}^{(-+)}_{Q,v}  \nonumber \\
&  = & \frac{1}{2\MQ} \bar{Q}_v (i\CDPNL)
e^{2iv\hspace{-0.15cm}\slash m_Q v\cdot x}
\Big(1-\frac{i\DSP}{2\MQ}\Big)^{-1} (i\DSCX ) Q_v   \\
& \equiv & \frac{1}{2\MQ} \bar{Q}_v (-i\DSCXL)
\Big(1-\frac{-i\oDSP}{2\MQ} \Big)^{-1}
e^{-2iv\hspace{-0.15cm}\slash m_Q v\cdot x}(i\CDPN) Q_v , \quad \
\qquad \ \qquad (2.34') \nonumber
\end{eqnarray}
with
\begin{eqnarray}
& & i\CDPN=i\DSP+\frac{1}{2\MQ}i\DSC \Big(1-\frac{i\DSP}{2\MQ}
\Big)^{-1} i\DSC ,
   \nonumber \\
 & & i\CDPNL = -i\oDSP +
 \frac{1}{2\MQ}   (-i\DSCXL)\Big(1-\frac{-i\oDSP}{2\MQ}\Big)^{-1} (-i\DSCXL) .
\end{eqnarray}
where we have expressed the effective Lagrangian for the
quark-antiquark coupling interactions in terms of two identical
formulations (2.34) and (2.34'). One can use either of them, which
mainly relies on the convenience for the relevant applications.
The factor $e^{\pm 2iv\hspace{-0.15cm}\slash m_Q v\cdot x}$ arises
from the opposite momentum shift for the effective heavy quark and
antiquark fields.

The above Lagrangian (eqs.(2.26-2.31) or eqs.(2.33-2.35)) form a
basic framework of LCQCD. Note that the resulting effective
Lagrangian is a complete one for the large component of effective
heavy quark and antiquark fields, i.e., $Q_v=Q_v^{(+)} +
Q_v^{(-)}$, we have only made the field redefinitions and
integrated out the small component of effective heavy quark and
antiquark fields, i.e., $R_v=R_v^{(+)} + R_v^{(-)}$.

\subsection{Lorentz Invariance of LCQCD}

Lorentz invariance is a fundamental requirement of a theory. As
the full QCD is established via Lorentz invariance, we shall check
how the Lorentz invariance holds in LCQCD which is directly
derived from the Lorentz invariant full QCD theory. Before
proceeding it would be useful to summarize the procedures for
constructing the LCQCD from QCD.
\begin{eqnarray}
{\cal L}^{QCD}_{Q} & & = \bar{Q}(i {\xslash D}-m_Q)Q \nonumber \\
& & \equiv (\bar{\hat{Q}}_v + \bar{R}_v)(i {\xslash D} -m_Q )(\hat{Q}_v+R_v) \nonumber \\
& & = \bar{\hat{Q}}_v [ 1+\overleftarrow{W}_v ( \frac{1}{m_Q} ) ]
(i {\xslash D}- m_Q)[1+W_v (\frac{1}{m_Q})]\hat{Q}_v \nonumber \\
& & = \bar{Q}_v e^{i \not{v} m_Q v \cdot x} [ 1+\overleftarrow{W}_v (\frac{1}{m_Q})]
( i {\xslash D}-m_Q) [ 1+W_v(\frac{1}{m_Q}) ] e^{-i \not{v} m_Q v \cdot x} Q_v \nonumber \\
& & = \bar{Q}_v i {\xslash v} v \cdot D Q_v
+{\cal L}^{QCD}_{Q,v}(\frac{1}{m_Q})\equiv {\cal L}^{QCD}_{Q,v}
\end{eqnarray}

From this procedure one sees that a crucial point is the
introduction of the unit time-like vector $v_{\mu}$ with $v^2=1$.
As this unit vector is arbitrary, one can introduce another unit
vector, say $v^{\prime}_{\mu}$ with ${v^{\prime}}^2 = 1$.
Repeating the same procedure, one then obtains an analogous
expression for the effective Lagrangian with just replacing $v$ by
$v^{\prime}$, namely
\begin{eqnarray}
{\cal L}^{QCD}_Q & & = \bar{Q} ( i {\xslash D} - m_Q ) Q \nonumber \\
& & \equiv \bar{Q}_v i {\xslash v} v \cdot D Q_v + {\cal
L}^{QCD}_{Q,v}
( \frac{1}{m_Q} ) \equiv {\cal L}^{QCD}_{Q,v} \nonumber \\
& & = \bar{Q}_{v^{\prime}} i {\xslash v^{\prime}} v^{\prime}
\cdot D Q_{v^{\prime}}+{\cal L}^{QCD}_{Q,v^{\prime}}
( \frac{1}{m_Q} ) \equiv {\cal L}^{QCD}_{Q,v^{\prime}}
\end{eqnarray}
which means that the effective Lagrangian in terms of the new
variable $Q_v$ with different unit time-like vector $v_{\mu}$
should be equivalent in describing the physics although the new
field $Q_v$ with different $v_{\mu}$ represents different field.
For convenience we refer to this feature as a super-equivalence of
LCQCD Lagrangian.

Keeping this point in mind, we now return to discuss the Lorentz
invariance of ${\cal L}^{QCD}_{Q,v}$. The full QCD theory is
invariant under the Lorentz transformation
\begin{eqnarray}
x^{\mu} \rightarrow {\Lambda}^{\mu}_{\nu} x^{\nu}, \qquad Q(x)
\rightarrow Q^{\prime}(x') = D(\Lambda)Q({\Lambda}^{-1}x)
\end{eqnarray}
with
\[ {\Lambda}^{\mu}_{\nu} = g^{\mu}_{\nu} + {\omega}^{\mu}_{\nu}, \qquad  D(\Lambda)
= e^{- \frac{i}{4} {\sigma}_{\mu \nu} {\omega}^{\mu \nu}} \]

From the definition of $Q_v$ , it is not difficult to show that
under the above transformation the $Q_v$ transforms as
\begin{eqnarray}
Q_v ( x ) & & \rightarrow Q^{\prime}_{v'} (x') \nonumber \\
 & & = e^{ i \not{v} m_Q v \cdot x } \frac{1+{\xslash v}}{2} D ( \Lambda ) Q^{(+)}
 ( {\Lambda}^{-1} x ) + e^{ i \not{v} m_Q v \cdot x } \frac{1- {\xslash v}}{2} D
 ( \Lambda ) Q^{ (-)} ( {\Lambda}^{-1} x ) \nonumber \\
& & = D(\Lambda) [D^{-1}(\Lambda) e^{ i \not{v} m_Q v \cdot x } D (\Lambda) ]
[D^{-1} (\Lambda) \frac{1+{\xslash v}}{2} D ( \Lambda )] Q^{(+)}( {\Lambda}^{-1} x ) \nonumber \\
& & + D(\Lambda) [D^{-1}(\Lambda) e^{ i \not{v} m_Q v \cdot x } D (\Lambda) ]
[D^{-1} (\Lambda)\frac{1- {\xslash v}}{2} D ( \Lambda )] Q^{ (-)} ( {\Lambda}^{-1} x )\nonumber \\
& & = D ( \Lambda) e^{ i \not{v^{\prime}} m_Q v^{\prime} \cdot x^{\prime} }
\frac{1+ {\xslash v^{\prime}}}{2} Q^{(+)} ( x^{\prime} ) + D (\Lambda)
e^{ i \not{v^{\prime}} m_Q v^{\prime} \cdot x^{\prime} }\frac{1- {\xslash v^{\prime}}}{2}
Q^{(-)}(x^{\prime}) \nonumber \\
& & = D ( \Lambda ) [e^{ i \not{v^{\prime}} m_Q v^{\prime} \cdot x^{\prime} }
\frac{1+ {\xslash v^{\prime}}}{2} Q^{(+)} ( x^{\prime} ) + e^{ i \not{v^{\prime}}
 m_Q v^{\prime} \cdot x^{\prime} }\frac{1- {\xslash v^{\prime}}}{2} Q^{(-)}(x^{\prime}) ] \nonumber \\
& & = D ( \Lambda ) Q_{v^{\prime}} ( x^{\prime} )
\end{eqnarray}
with \[ v^{\prime} = {\Lambda}^{-1} v, \qquad x^{\prime} =
{\Lambda}^{-1} x \]
 Here we have used the transformation property
\begin{eqnarray}
D^{-1} ( \Lambda ) {\xslash v } D ( \Lambda ) = {\xslash v^{\prime}}
\end{eqnarray}
which indicates that under the Lorentz transformation the vector
$v_{\mu}$ should transform like the coordinate $x_{\mu}$. The
Lagrangian ${\cal L}^{QCD}_{Q,v}$ becomes, under the Lorentz
transformation, to be
\begin{eqnarray}
{\cal L}^{QCD}_{Q,v} \rightarrow {\cal L}^{QCD}_{Q,v^{\prime}} =
\bar{Q}_{v^{\prime}} i {\xslash v^{\prime}} v^{\prime} \cdot D
Q_{v^{\prime}} + {\cal L}^{QCD}_{Q,v^{\prime}} ( \frac{1}{m_Q} )
\end{eqnarray}
The Lorentz invariance is recovered by the super-equivalence of
LCQCD Lagrangian as shown explicitly above.

   Note that when the effective Lagrangian is truncated in the
expansion of $1/m_Q$ to only finite terms, the Lorentz invariance
will be broken down.

\subsection{Quark-antiquark Couplings and Their Decouple
Conditions}

Here we would like to address two important issues:  Firstly, in
the effective lagrangian of LCQCD the quark-antiquark coupling
terms appear naturally as long as one treats the quark and
antiquark fields on the same footing in a symmetric way. This is
actually the basic feature of quantum field theory. It is also
explicitly seen that in the infinite mass limit, the quark and
antiquark fields decouple each other, namely quark-antiquark
coupling terms approach to vanish. In general, we will show below
that such quark-antiquark coupling terms will provide important
contributions starting from the order of $1/m_Q$. Therefore in
order to calculate correctly the $1/m_Q$ corrections arising from
the finite mass of heavy quark, one should take into account the
contributions from the terms ${\cal L}^{(\pm\mp)}_{Q,v}$ and
${\cal L}^{(--)}_{Q,v}$ for the processes involving the heavy
quark $Q^{(+)}_v$ or from the terms ${\cal L}^{(\pm\mp)}_{Q,v}$
and ${\cal L}^{(++)}_{Q,v}$ for the ones involving the heavy
antiquark $Q^{(-)}_v$ . This is because through virtual antiquark
or quark field exchanges they will produce $1/m_Q$ corrections via
quantum effects. Only in the infinite mass limit, the
quark-antiquark coupling terms decouple in the sense of effective
quantum field theory, namely
\begin{eqnarray}
\hat{{\cal L}}_{Q,v}  \to \hat{{\cal L}}^{(m_Q\to \infty)}_{Q,v}
&=& \QVHPB [i\dsp -\MQ] \QVHP  +  \QVHFB [i\dsp -\MQ] \QVHF
\end{eqnarray}

Secondly, one may not use the whole Dirac equation of motion
either for quark component or antiquark component. Otherwise,
either the quark field or antiquark field will decouple from
effective Lagrangian, which leads the effective heavy quark and
antiquark fields to be treated not on the same footing via a
symmetric way. To be explicit, it is easily seen that if one
further uses the Dirac equation of motion for the effective heavy
antiquark field, namely,
\begin{eqnarray}
(i\dsp - m_Q) \hat{Q}_v^{(-)}   + i\DSC R_v^{(-)} = 0,\quad
\mbox{i.e.} \quad (i\CDP -\MQ ) \hat{Q}_v^{(-)}  = 0
\end{eqnarray}
then the heavy quark effective Lagrangian is reduced to be
\begin{equation}
        \hat{{\cal L}}_{Q,v} \to \hat{{\cal L}}^{(++)}_{Q,v} = \QVHB^{(+)}[i\CDP
        -\MQ]\QVH^{(+)}
     \end{equation}
which recovers the widely used heavy quark effective Lagrangian
for effective heavy quark field $\QVH^{(+)}$. Similarly, if
further adopting the Dirac equation of motion for effective heavy
quark field
\begin{eqnarray}
(i\dsp - m_Q) \hat{Q}_v^{(+)}   + i\DSC R_v^{(+)} = 0, \quad
\mbox{i.e.} \quad (i\CDP -\MQ ) \hat{Q}_v^{(+)}  = 0
\end{eqnarray}
one arrives at the heavy quark effective Lagrangian
\begin{equation}
        \hat{{\cal L}}_{Q,v} \to \hat{{\cal L}}^{(--)}_{Q,v} = \QVHB^{(-)}[i\CDP
        -\MQ]\QVH^{(-)}
     \end{equation}
which reproduces the widely used heavy quark effective Lagrangian
for effective heavy antiquark field $\QVH^{(-)}$.

 From the above analyzes, it is not difficult to
find out an implicit assumption made in the widely used heavy
quark effective theory which describes either quark field or
antiquark field and contains no quark-antiquark coupling terms. It
also implies that the assumption of particle and antiparticle
numbers being separately conserved made in the derivation of
widely used heavy quark effective Lagrangian is somehow equivalent
to the case that the Dirac equation of motion has been imposed for
either antiquark field or quark field.

We can now conclude that in the case of $ |{\bf p}| << E + \MQ$
but without imposing Dirac equation of motion for either quark
field or antiquark field, QCD shall be described by a large
component QCD (LCQCD) with containing both large component quark
and antiquark fields in the heavy quark effective lagrangian.

\subsection{ HQEFT from LCQCD}

For any practical application of LCQCD to heavy quark systems, one
may further expand the LCQCD Lagrangian into a series in powers of
the inverse heavy quark mass.  In order to correctly perform such
an expansion, one must keep an eye on the physical conditions
relating to the concrete physical systems. In general, the
``longitudinal" and ``transverse" residual momenta of heavy quarks
can be at different powers for different physical systems or
processes. Correspondingly, the operator $\dsp$ and $\DSC$ may be
treated as operators in different power counting in the $1/m_Q$
expansion. We will consider below two interesting cases which are
corresponding to two typical conditions.

  One interesting case is for the heavy quark being nearly on-mass
shell. That may happen for heavy quarks within the heavy quarkonia
system. This implies the operator relation for the $1/m_Q$
expansion
   \begin{equation}
   \label{2.46}
   i\DSP  = - \frac{(i\DSC)^2}{2\MQ}  + O\Big(\frac{1}{m_Q^2} \Big) .
   \end{equation}
With this condition, it is not difficult to see that the
quark-antiquark coupling terms ${\cal L}^{(2)}_{Q,v}$ are
suppressed by the higher order terms of $1/m_Q$. Thus the
quark-quark and antiquark-antiquark coupling terms ${\cal
L}^{(1)}_{Q,v}$ become dominant. This case may be dealt with like
a nonrelativistic QCD (NRQCD). Such a case was also discussed in
ref.\cite{BS} and applied to heavy quark pair creations near the
threshold.

 Another interesting case is for heavy quark being slightly
off-mass shell. The typical case is considered for the heavy
hadron containing a single heavy quark. The magnitude of the
off-mass shell is usually thought to be at the order of binding
energy with the off-mass shell condition
\begin{eqnarray}
{p^2 - m^2_Q \over 2 m_Q} \sim \bar{\Lambda},\nonumber
\end{eqnarray}
Taking $p= m_Q v + k $ with $v= (1, 0, 0, 0)$, one has
\begin{eqnarray}
k^0 \sim \vert { \bf k}\vert \sim \bar{\Lambda} , \qquad
\mbox{i.e.} \quad  v\cdot k \sim |\KC| \sim  \bar{\Lambda}
\nonumber
\end{eqnarray}
Here $\bar{\Lambda} \sim 2\Lambda_{QCD} \sim 500\sim 600$ MeV is
the typical binding energy of heavy hadrons. In the operator
basis, the above condition is corresponding to:
\begin{eqnarray}
\label{2.47}
 \langle i\DP \rangle \sim \langle i\DC \rangle \sim
\bar{\Lambda},
\end{eqnarray}
which implies that the operators corresponding to ``longitudinal"
and ``transverse" residual momenta are taken to be at the same
order of power counting in the $1/m_Q$ expansion. In this case,
the quark-antiquark coupling terms  ${\cal L}^{(2)}_{Q,v}$ are at
the order of $O\Big(1/m_Q \Big)$. In this sense, LCQCD is treated
as a heavy quark effective field theory (HQEFT) via $1/m_Q$
expansion.

To see explicitly the effects of antiquark field, i.e., the
contributions from quark-antiquark coupling terms, we may
integrate out the effective heavy antiquark fields, which is
equivalent, at the tree level, to adopt the following relations
between quark and antiquark field to eliminate the antiquark field
in the effective Lagrangian of LCQCD
\newcommand{\LEFT}[1]{\overleftarrow{\hat{#1}}}
\begin{eqnarray}
&&(\CDP -m_Q )\QVHF + [i\DSC (\MQ-i\DSP)^{-1}(i\CDP -\MQ)]\QVHZ =0 , \\
&& \QVHFB (\CDPL - m_Q) + \QVHZB [i\DSCXL
    (\MQ-i\oDSP)^{-1} (i\CDPL -\MQ)] = 0 ,
\end{eqnarray}
or
\begin{eqnarray}
 \label{eq:new1a}
&& \QVHF=-(\CDP -m_Q )^{-1} [i\DSC (\MQ-i\DSP)^{-1}(i\CDP -\MQ)]\QVHZ ,\\
\label{eq:new1b} && \QVHFB=-\QVHZB [i\DSCXL (\MQ-i\oDSP) (i\CDPL
-\MQ)] (\CDPL - m_Q)^{-1}.
\end{eqnarray}
 The resulting effective Lagrangian for quark field has the
 following form
\begin{eqnarray}
\label{eq:20} {\cal L}^{(++)}_{eff}&=& {\cal L}^{(++)}_{Q,v} +
\tilde{\cal L}^{(++)}_{Q,v},
\end{eqnarray}
The second part $\tilde{\cal L}^{(++)}_{Q,v}$ comes from the
contributions of integrating out the effective heavy quark
antiquark field. Its explicit form is found to be
\begin{eqnarray}
\tilde{\cal L}^{(++)}_{Q,v} &=& \langle {\cal L}^{(--)}_{Q,v}+
    {\cal L}^{(+-)}_{Q,v}+{\cal L}^{(-+)}_{Q,v}
    \rangle \vert_{\hat{Q}^{(-)}_v \to \hat{Q}^{(+)}_v} \\
&=&-\QVHZB
i\DSC(\MQ-i\DSP)^{-1}i\DSC(\MQ-i\DSP)^{-1}(i\CDP-\MQ)\QVHZ
\nonumber
\end{eqnarray}
After removing the large mass term in the above effective
Lagrangian, we then obtain the following compact form
 \begin{equation}
 \label{complarg}
  {\cal L}^{(++)}_{eff} = \qpvb i\CDPN {1\over i\DSP} i\CDPN \qpv
 \end{equation}
which may be rewritten into the following explicit form
 \begin{eqnarray}\label{Leff++}
{\cal L}^{(++)}_{eff} &=& \qpvb i\DSP \qpv \nonumber \\
 & & + 2\cdot
\frac{1}{2\MQ} \qpvb (i\DSC) \Big(1 - {i\DSP\over 2 \MQ}\Big)^{-1}
(i\DSC) \qpv \nonumber \\
&&+ \frac{1}{4\MQ^2}\qpvb(i\DSC) \Big(1 - {i\DSP\over 2
\MQ}\Big)^{-1} (i\DSC) \nonumber
\\
&&\times \frac{1}{i\DSP}(i\DSC) \Big(1 - {i\DSP\over 2
\MQ}\Big)^{-1} (i\DSC)\qpv .
\end{eqnarray}
It must be noted that the power counting order of $1/m_Q$
expansion for the third term in above equation will depend on the
order of power counting for the operators $\DSP$ and $\DSC$.

We now return to discuss the two interesting cases considered
above. For the first case with the condition (\ref{2.46}), the
third term in (\ref{Leff++}) is actually at the same order of
power counting as the second term in (\ref{Leff++}). This is
because the ``longitudinal" operator $\DSP$ is suppressed by
$1/m_Q$ in comparison with the ``transverse" operator $\DSC$.
Substituting to the condition (\ref{2.46}) to the third term of
the above effective Lagrangian (\ref{Leff++}), we then arrive at
the following effective Lagrangian expanding in terms of the
$1/m_Q$
 \begin{eqnarray}\label{Neff++}
{\cal L}^{(++)}_{eff} & = &
 \qpvb \Big\{i\DSP+\frac{1}{2\MQ}(i\DSC)^2 \Big\}\qpv  + O\Big(\frac{1}{m_Q^2}
  \Big).
\end{eqnarray}
Taking $v = (1, 0, 0, 0)$ and $A_0 = 0$, one has
\begin{eqnarray}
{\cal L}^{(++)}_{eff}
 = \qpvb \Big\{i {\partial \over \partial t} + {{ \bf \nabla}^2 \over 2m_Q}
  \Big\}\qpv + O\Big({1\over m^2_Q} \Big).
\end{eqnarray}
which recovers the effective Lagrangian in NRQCD.

For the second case with the condition (\ref{2.47}), the third
term in (\ref{Leff++}) can now be treated as $1/m_Q^2$ order and
the second term in (\ref{Leff++}) as $1/m_Q$ order. Thus the
Lagrangian (\ref{Leff++}) can be expanded in terms of $1/m_Q$ into
the following form
    \begin{eqnarray}
    \label{Heff++}
      {\cal L}^{(++)}_{eff}= {\cal L}^{(0)}_{eff}+{\cal L}^{(1/\MQ)}_{eff}
    \end{eqnarray}
with
    \begin{eqnarray}
    \label{Heff++0}
     {\cal L}^{(0)}_{eff} & = & \qpvb(i\DSP)\qpv ,  \\
    \label{Heff++m}
     {\cal L}^{(1/\MQ)}_{eff} & = & 2\ \frac{1}{2\MQ}\qpvb(i\DSC)^2\qpv +
   2\ \frac{1}{4\MQ^2}\qpvb i\DSC (i\DSP)i\DSC \qpv \nonumber \\
           && + \frac{1}{4\MQ^2}\qpvb (i\DSC)^2 \frac{1}{i\DSP}(i\DSC)^2 \qpv
           +O\Big(\frac{1}{\MQ^3} \Big) ,
    \end{eqnarray}
where ${\cal L}^{(0)}_{eff}$ is the leading term and possesses the
spin-flavor symmetry. The second part ${\cal L}^{(1/\MQ)}_{eff}$
contains the remaining terms which are the spin-flavor symmetry
breaking ones suppressed by $1/m_{Q}$. Here we have used the
condition for the operators: $ (i\DSC)^2/2m_Q  \ll i\DSP $ which
always holds for the typical case that $ \langle i\DP \rangle \sim
\langle i\DC \rangle \sim \bar{\Lambda} \ll m_Q$  in the heavy
quark expansion. In this case, the third term in the rhs. of
Eq.(\ref{Heff++m}) is now regarded as at the order of $1/m^2_Q$
and the `nonlocal' factor $1/i\DSP$ actually arises as a general
propagator due to the virtual effective antiquark exchange in
HQEFT (see below). The first term in the rhs. of
Eq.(\ref{Heff++m}) represents the total $1/m_Q$ order correction
to the leading order Lagrangian (\ref{Heff++0}). One may notice
that there is a factor of two in the first two terms of
Eq.(\ref{Heff++m}), which can now be well understood from the
$1/m_Q$ expansion in HQEFT. It clearly arises from the antiquark
contributions when the heavy quark is considered to be slightly
off-mass shell with $\langle i\DP \rangle \sim \langle i\DC
\rangle \sim \bar{\Lambda}$, which satisfies the condition in the
momentum space that $v\cdot k\sim \bar{\Lambda} > |\KC^2|/2m_Q$
(i.e., $|\KC| < \sqrt{2\bar{\Lambda}m_Q}$).

  It is seen that the treatment of HQEFT should differ from the one of
NRQCD. Besides the explicit difference for the factor of two in
the $1/m_Q$ and $1/m_Q^2$ terms, the propagators in the treatment
of effective field theories for NRQCD and HQEFT must also be
different. The propagator in NRQCD should take the following form
 \begin{eqnarray}
    \frac{i}{{\xslash v} v\cdot k + \frac{1}{2\MQ}(\KSC)^2}
 \end{eqnarray}
as the two terms in the denominator are compatible for the nearly
on-mass shell condition. Whereas the propagator in HQEFT is mainly
governed by the ``longitudinal" part and will be shown late on to
have the following simple form
\begin{eqnarray}
    \frac{i}{{\xslash v} v\cdot k}
 \end{eqnarray}

It is necessary to address that only with the above considerations
one is able to obtain consistent effective field theories within
the framework of LCQCD derived directly from full QCD. Obviously,
the above analyzes also make HQEFT different from the treatment of
the widely used heavy quark effective theory in which the quark
and antiquark fields are not treated on the same footing in a
symmetric way in the widely used heavy quark effective theory.

We now arrive at the conclusion that when applying LCQCD to the
physical system in which the single heavy quark within a hadron is
slightly off-mass shell with the typical energy region that
$\langle i\DP \rangle \sim \langle i\DC \rangle \sim
\bar{\Lambda}$, namely $v\cdot k\sim \bar{\Lambda} \gg
|\KC^2|/2m_Q$ (i.e., $|\KC| \ll \sqrt{2\bar{\Lambda}m_Q}$), only
HQEFT described above becomes appropriate and consistent. As the
effective heavy quark and antiquark fields in such a kind of HQEFT
are treated on the same footing in a fully symmetric way, which
also fits to the spirit of quantum field theory as the original
QCD does, it then allows us to establish a complete theoretical
framework of HQEFT in the sense of effective quantum field theory.
This comes to the main task in the following sections.

 Note that one should avoid some misleading questions raised from the
 usual heavy quark effective theory for the quark-antiquark coupling, such as
{\it no such terms could appear in the effective Lagrangian because
the soft gluon-quark-antiquark vertex does not conserve momentum}.
As we have shown in the subsection (2.1) that our effective
Lagrangian is formally derived from the full QCD by just integrating
out the small components of heavy quark fields, all terms must exist
in the effective Lagrangian. The momentum conservation only
constraints that a soft gluon cannot create a heavy quark-antiquark
pair, which can explicitly be seen from the factor $exp(-i2m_Q. x)$
in the quark-antiquark coupling terms when introducing the residual
momentum, but it doesn't mean that such quark-antiquark coupling
terms cannot appear in the effective Lagrangian as the soft
gluon-quark-antiquark vertex can always be momentum conserved as
long as the quark and antiquark fields are slightly off-mall shell
with one of them is virtual. In this case, both "quark" and
"anti-quark" fields actually carry either positive energy or
negative energy. Nevertheless, when the virtual "anti-quark" field
carries positive energy, it is suppressed by $1/m_Q$. Similarly,
when virtual "quark" field carries negative energy, it is also
suppressed by $1/m_Q$. In fact, even in the full QCD soft gluon
cannot create the quark-antiquark pairs but the quark-antiquark
coupling term remains there. It is easily shown that the
quark-antiquark coupling terms with soft gluon can have
contributions to the quark-antiquark scattering via t-channel, it
can also provide important effects for the quark $\to$ quark
transition via antiquark field mediation or antiquark $\to$
antiquark transition via quark field mediation. The latter is
interesting for studying heavy quark decays in the HQEFT. This is
what we are going to demonstrate in details in this paper, it is
actually the natural features of quantum field theory.

\section{ Quantization of HQEFT }

\subsection{Quantum Generators of Poincare Group }

 Let us first consider the case of a free field theory for a single quark at
infinite mass limit. The effective Lagrangian for this case is
simple
\begin{eqnarray}
{\cal L}_{Q,v}^{(00)} = \bar{Q}_v i {\xslash v } v \cdot \partial
Q_v
\end{eqnarray}
The equation of motion reads
\begin{eqnarray}
i {\xslash v} v \cdot \partial Q_v = 0
\end{eqnarray}
Since the field $Q_v (x)$ and $\bar{Q}_v (x)$ concern the
additional arbitrary unit time-like vector $v_{\mu}$, the
canonical quantization is not quite clear for it, we cannot
directly follow the procedure of canonical quantization and will
adopt an alternative procedure. Firstly it is assumed that $Q_v (
x ) $ and $\bar{Q}_v ( x ) $ are quantized operators acting in an
Hilbert space, we then look for the conditions under which
Poincare group generators constructed from Noether's theorem will
lead the fields to transform according to the required
transformation laws. Since these generators and also all
observable quantities are expressed in terms of the dynamical
variables $Q_v ( x )$ and $\bar{Q}_v (x)$, we shall verify that
these observables do commute at space-like separations and the
theory is ensured to be covariant.

Consider an infinitesimal $x$-dependent translation $x \rightarrow
x + a ( x ) $, thus
\begin{eqnarray}
& & \delta Q_v = \delta a^{\mu} {\partial}_{\mu} Q_v, \qquad
\delta \bar{Q}_v =
\bar{Q}_v \overleftarrow{{\partial}_{\mu}} \delta a^{\mu}\nonumber \\
& & \delta [ {\partial}_{\mu} Q_v ] = \delta a^{\nu}
{\partial}_{\nu} {\partial}_{\mu} Q_v + {\partial}_{\mu} [ \delta
a^{\nu}] {\partial}_{\nu} Q_v
\end{eqnarray}
The corresponding variation of the action after an integration by
parts is
\begin{eqnarray}
\delta I_v = \int {\partial}_{\mu} [ g^{\mu \nu} \bar{Q}_v i
{\xslash v} v \cdot \partial Q_v - \bar{Q}_v i {\xslash v} v^{\mu}
{\partial}^{\nu} Q_v ] \delta a_{\nu}
\end{eqnarray}
From the vanishing of $\delta I_v$ for arbitrary $ \delta
a^{\nu}(x)$, and using the equation of motion, we deduce that the
energy momentum flow is described by the canonical tensor
\begin{eqnarray}
{\Theta}^{\mu \nu}_v = \bar{Q}_v i {\xslash v} v^{\mu}
{\partial}^{\nu} Q_v
\end{eqnarray}
which satisfies the conservation law
\begin{eqnarray}
{\partial}_{\mu} {\Theta}^{\mu \nu}_v =0
\end{eqnarray}

Consider now an infinitesimal homogeneous Lorentz transformation
under which we have
\begin{eqnarray}
Q_v (x) \rightarrow Q^{\prime}_{v^{\prime}} ( x^{\prime} ) = D (
\Lambda ) Q_{v^{\prime}} ( {\Lambda}^{-1} x) = Q_{v^{\prime}} ( x
) + \delta Q_{v^{\prime}} ( x )
\end{eqnarray}
with
\[ \delta Q_{v^{\prime}} ( x ) = - \frac{i}{2} \delta \omega_{\alpha \beta}
L^{\alpha \beta} Q_{v^{\prime}} ( x ), \qquad L^{\alpha \beta} =
\frac{{\sigma}^{\alpha \beta}}{2} + i (
x^{\alpha}{\partial}^{\beta} - x^{\beta}{\partial}^{\alpha} ) \]
and
\begin{eqnarray}
v^{\prime} = {\Lambda}^{-1} v =v + \delta v, \qquad  {\delta
{\xslash v}^{\prime}} = \frac{i}{2} \delta {\omega}_{\alpha \beta
} [ i ( {v^{\prime}}^{\alpha} {\gamma}^{\beta} -
{v^{\prime}}^{\beta} {\gamma}^{\alpha})]
\end{eqnarray}
The corresponding variation of the action is
\begin{eqnarray}
I_v \rightarrow I^{\prime}_v =& & \int d^4 x [
\bar{Q}_{v^{\prime}} i {\xslash v^{\prime}} v^{\prime} \cdot
\partial Q_{v^{\prime}} + \delta \bar{Q}_{v^{\prime}} i {\xslash
v^{\prime}} v^{\prime}
\cdot \partial Q_{v^{\prime}} \nonumber \\
& &+ \bar{Q}_{v^{\prime}} i {\xslash v^{\prime}} v^{\prime} \cdot
\partial \delta Q_{v^{\prime}}
+ \bar{Q}_{v^{\prime}} i \delta {\xslash v^{\prime}}
v^{\prime} \cdot \partial Q_{v^{\prime}} \nonumber \\
& &+\bar{Q}_{v^{\prime}} i {\xslash v^{\prime}} \delta (
v^{\prime} \cdot \partial ) Q_{v^{\prime}}]
\end{eqnarray}
Thus we find that as a variation around the stationary point
\begin{eqnarray*}
\delta I_{v^{\prime}} = - \int d^4 x ({\partial}_{\mu} J^{\mu ,
\alpha \beta }_{v^{\prime}} ) \frac{{\omega}_{\alpha \beta}}{2} =
0
\end{eqnarray*}
with
\begin{eqnarray}
 J^{\mu , \alpha \beta }_{v^{\prime}} = {v^{\prime}}^{\mu}
\bar{Q}_{v^{\prime}} {\xslash v^{\prime}} L^{\alpha \beta}
Q_{v^{\prime}} = x^{\alpha} {\Theta}^{\mu \beta}_{v^{\prime}} -
x^{\beta} {\Theta}^{\mu \alpha}_{v^{\prime}} + {v^{\prime}}^{\mu}
\bar{Q}_{v^{\prime}} {\xslash v^{\prime}} \frac{{\sigma}^{\alpha
\beta}}{2} Q_{v^{\prime}}
\end{eqnarray}
 which has an orbital and a spin part, and satisfies the
 conservation law
\begin{eqnarray}
{\partial}_{\mu} J ^{\mu, \alpha \beta }_{v^{\prime}} =0
\end{eqnarray}
One can also directly check this conservation law.

Thus the quantum generators of the Poincare group would be yielded
by the space integrals of ${\Theta}^{0 \mu}_v $ and $J^{0, \alpha
\beta}_{v^{\prime}}$
\begin{eqnarray}
K^{\mu}_v = \int d^3 x {\Theta}^{0 \mu}_v, \qquad J^{\mu
\nu}_{v^{\prime}} = \int d^3 x J^{0, \mu \nu}_{v^{\prime}}
\end{eqnarray}
which will be used to find a commutation prescription for the
effective fields $Q_v$ and $\bar{Q}_v$.

\subsection{Anticommutations and Georgi's Velocity Super-selection Rule}

To establish the commutation prescription for the effective
fields, we expand the operators $Q_v$ and $\bar{Q}_v$ in terms of
the $c$-number plane wave solutions of the equation of motion
eq.(3.2) with operator-valued amplitudes $b$, $b^{\dagger}$, $d$,
$d^{\dagger}$
\begin{eqnarray}
& & Q_v ( x ) = \int \frac{d^3 k}{{(2 \pi)}^3 v^0}
\sum_s [ b_v ( k, s ) u(v,s) e^{- i k \cdot x} +
d^{\dagger}_v ( k, s ) v(v,s) e^{i k \cdot x } ] \nonumber \\
& & \bar{Q}_v ( x ) = \int \frac{d^3 k}{{(2 \pi)}^3 v^0} \sum_s [
b^{\dagger}_v ( k,s) \bar{u}(v, s) e^{i k \cdot x} + d_v ( k,s)
\bar{v} (v,s) e^{-i k \cdot x} ]
\end{eqnarray}
where spinors $u$ and $v$ satisfy
\begin{eqnarray}
{\xslash v} u (v,s) =u(v,s), \qquad {\xslash v} v ( v, s ) = -v
(v, s)
\end{eqnarray}
and normalization conditions
\begin{eqnarray}
& & \bar{u} (v, s) u(v, s^{\prime})= {\delta}_{s s^{\prime}},
\qquad \bar{v}(v, s)v(v, s^{\prime})
= - \delta_{s s^{\prime}}\nonumber \\
& & \bar{u}(v,s)v(v,s^{\prime}) =\bar{v}(v, s)u (v, s^{\prime}) =0
\end{eqnarray}
Here $k^0$ is defined through $ k \cdot v =0$ due to the equation
of motion. The operators $b$ and $d$ must satisfy commutation
rules such that the Poincare group generators transform the fields
according to the required transformation laws as indicated from
above subsection.
\begin{eqnarray}
& & Q_v ( x + a ) = e^{i K_v \cdot a} Q_v ( x ) e^{-i K_v \cdot a}\nonumber \\
& & Q^{\prime}_{v^{\prime}} ( x^{\prime} ) = e^{i J^{\mu
\nu}_{v^{\prime}} {\omega}_{\mu \nu}/2 } Q_v ( x ) e^{- i J^{\mu
\nu}_{v^{\prime}} {\omega}_{\mu \nu}/2}
\end{eqnarray}
with $ v^{\prime} = {\Lambda}^{-1} v$ and
$x^{\prime}={\Lambda}^{-1} x $. In the differential form, they can
be rewritten as
\begin{eqnarray}
& & {\partial}^{\mu} Q_v ( x ) = i [ K^{\mu}_v, Q_v ( x ) ],
\qquad
{\partial}^{\mu} \bar{Q}_v ( x ) = i [ K^{\mu}_v, \bar{Q}_v ( x ) ]\\
& & L^{\mu \nu} Q_{v^{\prime}} ( x ) = - [ J^{\mu
\nu}_{v^{\prime}}, Q_v ( x ) ], \qquad \bar{Q}_{v^{\prime}} ( x )
\overleftarrow L^{\mu \nu} = -[J^{\mu \nu}_{v^{\prime}}, \bar{Q}_v
( x ) ]
\end{eqnarray}

Let us first consider the requirement of translational invariance.
Expressing $K^{\mu}_v$ with the decomposition eqs.(3.5), (3.12)
and (3.13)
\begin{eqnarray}
K^{\mu}_v = \int d^3 x {\Theta}^{0 \mu}_v = \int \frac{d^3 x}{{(2
\pi)}^3 v^0} k^{\mu} \sum_s [ b^{\dagger}_v ( k, s ) b_v (k, s) -
d_v (k, s)d^{\dagger}_v (k, s ) ]
\end{eqnarray}
Note that one has to subtract the vacuum contribution in this
expression. If the vacuum is defined in such a way that $b_v (k ,s
)|0>= d_v (k, s )|0> =0 $, we see that the operators $b$ and $d$
have to quantize according to anticommutators rather than
commutators, otherwise $b$ particles and $d$ particles contribute
with opposite signs to the energy and the theory would not admit a
stable ground state.

To obtain explicitly the commutation rule, we express the
differential form of the translational invariance in eq.(3.17) in
the momentum space. It is satisfied provided
\begin{eqnarray}
& & [K^{\mu}_v, b_v (k,s)] = - k^{\mu} b_v ( k, s), \qquad
[K^{\mu}_v, d_v (k,s)] = -k^{\mu} d_v (k, s)\nonumber \\
& & [K^{\mu}_v, b^{\dagger}_v ( k, s) ] = k^{\mu} b^{\dagger}_v (
k, s ), \qquad [K^{\mu}_v, d^{\dagger}_v ( k, s ) ] = k^{\mu}
d^{\dagger}_v (k, s )
\end{eqnarray}
Using the explicit form of $K^{\mu}_v$, one can find that to
ensure the correct interpretation of energy and momentum the
operators should satisfy the anticommutation relations
\begin{eqnarray}
& & \{b_v ( k^{\prime}, s^{\prime} ), b^{\dagger}_v ( k , s ) \}
= {(2 \pi)}^3 v^0 {\delta}^3 ( \overrightarrow{k} -
{\overrightarrow{k}}^{\prime} ) {\delta}_{s,s^{\prime}}\nonumber \\
& & \{d_v ( k^{\prime}, s^{\prime} ), d^{\dagger}_v ( k, s ) \} =
{(2 \pi)}^3 v^0 {\delta}^3 ( \overrightarrow{k} -
{\overrightarrow{k}}^{\prime} ){\delta}_{s,s^{\prime}}
\end{eqnarray}
and all other anticommutators vanish.

We now turn to the requirement of the Lorentz transformation
invariance. Expressing also $J^{\mu \nu}_{v^{\prime}}$ with the
decomposition eqs. (3.10), (3.12) and (3.13)
\begin{eqnarray}
J^{\mu \nu}_{v^{\prime}} = \int d^3 x J^{0, \mu \nu}_{v^{\prime}}
=& & \int \frac{d^3 k}{{(2 \pi)}^3 {v^{\prime}}^0} \sum_s
\sum_{s^{\prime}}[ b^{\dagger}_{v^{\prime}} ( k, s) b_{v^{\prime}}
( k, s^{\prime} ) \bar{u} ( v^{\prime}, s ) {\xslash v^{\prime}}
L^{\mu \nu} u ( v^{\prime}, s^{\prime} )\nonumber \\
& & + b^{\dagger}_{v^{\prime}} ( k, s ) d_{v^{\prime}} ( -k ,
s^{\prime} ) \bar{u} ( v^{\prime}, s )
{\xslash v^{\prime}} L^{\mu \nu} v ( v^{\prime}, s^{\prime} ) \nonumber \\
& & + d_{v^{\prime}} ( k, s ) b_{v^{\prime}} ( - k , s^{\prime} )
\bar{v} ( v^{\prime}, s )
{\xslash v^{\prime}} L^{\mu \nu} u ( v^{\prime}, s^{\prime} ) \nonumber \\
& & + d_{v^{\prime}} ( k, s ) d^{\dagger}_{v^{\prime}} ( k,
s^{\prime} ) \bar{v} ( v^{\prime}, s) {\xslash v^{\prime}} L^{\mu
\nu} v ( v^{\prime}, s^{\prime} ) ]
\end{eqnarray}
From eq.(3.18), it is seen that the commutations involve operators
with different velocities. By noticing the following facts
\begin{eqnarray}
& & \bar{u} ( v, s ) {\xslash v } = \bar{u} (v, s), \qquad \bar{v}
( v, s ) {\xslash v}
= - \bar{v} ( v, s )\nonumber \\
& & \sum_s u ( v, s ) \bar{u} ( v, s ) = \frac{1+{\xslash v }}{2}
, \qquad \sum_s v(v, s) \bar{v} ( v, s )
 = - \frac{1-{\xslash v }}{2}
\end{eqnarray}
as well as the indication of commutation rules obtained from the
translational invariance , it is not difficult to find that the
anticommutators of operators with different velocity must satisfy
\begin{eqnarray}
& & \{ b_{v^{\prime}} ( k^{\prime}, s^{\prime} ), b^{\dagger}_v (
k, s ) \} = {(2 \pi )}^3 v^0 {\delta}^3
( \overrightarrow{k} -\overrightarrow{k}^{\prime} )
\delta_{s s^{\prime}} \delta_{v v^{\prime}}\nonumber \\
& & \{ d_{v^{\prime}} ( k^{\prime}, s^{\prime} ), d^{\dagger}_v (
k, s ) \} = {(2 \pi )}^3 v^0 {\delta}^3 ( \overrightarrow{k}
-\overrightarrow{k}^{\prime} )\delta_{s s^{\prime}} \delta_{v
v^{\prime}}
\end{eqnarray}
and all other anticommutators vanish. Where $\delta_{v
v^{\prime}}$ is a Kronecker delta function, i.e. $\delta_{v
v^{\prime}} =1$ if $v = v^{\prime}$ and $\delta_{v v^{\prime}}=0$
if $v \not= v^{\prime}$. This may be mentioned, as it was first
imposed by Georgi\cite{HG}, to be Georgi's velocity
super-selection rule which is in principle a consequence of
Lorentz invariance.

From the above commutation rules of operators, we arrive at the
commutation rule for the effective fields
\begin{eqnarray}
\{\Pi^\alpha_v ( x ), Q^\beta_{v^{\prime}} ( x^{\prime} ) \}
|_{x_0 = x^{\prime}_0} = i \delta_{\alpha \beta} \delta_{v
v^{\prime}} \delta^3 ( \overrightarrow{x} -
\overrightarrow{x}^\prime )
\end{eqnarray}
where $\Pi_v$ is the momentum conjugate to the field $Q_v$ and
given by
\begin{eqnarray}
\Pi_v = \frac{\delta L }{\delta ( \partial_0 Q_v )} = \bar{Q}_v i
{\xslash v} v^0
\end{eqnarray}
This commutation rule of effective fields shows how the HQEFT of
LCQCD is quantized canonically.

We can now use the Wick products to define correctly the total
energy momentum since when reordering the creation operators to
the left of the annihilation ones, a sign corresponding to the
parity of the permutation must be introduced, namely
\begin{eqnarray}
K^\mu_v & & = \int \frac{d^3 k}{{(2 \pi)}^3 v^0} k^\mu \sum_s :
b^\dagger_v ( k, s ) b_v ( k, s )
- d_v (k, s) d^\dagger_v ( k, s) :\nonumber \\
& & = \int \frac{d^3 k}{{(2 \pi)}^3 v^0} k^\mu \sum_s [
b^\dagger_v ( k, s ) b_v (k, s ) + d^\dagger_v ( k, s ) d_v ( k,s
) ]
\end{eqnarray}
which leads to a sum of positive contributions to the energy of a
quantum state. We are now in the position to construct the
corresponding Hilbert and Fock space for HQEFT of LCQCD.

\subsection{Hilbert and Fock Space of HQEFT}

Let us first consider single-particle states. The necessary
smearing in momentum space is implicitly understood. For a given
four-velocity $v_\mu$ and four-momentum $k_\mu$ there is a
fourfold degeneracy. Denote the corresponding states $|I> ( I = 1,
2 , 3, 4 ) $ with
\begin{eqnarray}
& & |1> = b^\dagger_v ( k, + ) |0>, \qquad |2> = b^\dagger_v ( k, - ) |0> \nonumber \\
& & |3> = d^\dagger_v ( k, + ) |0>, \qquad |4> = d^\dagger_v ( k,
- ) |0>
\end{eqnarray}
They satisfy
\begin{eqnarray}
K^\mu_v |I > = k^\mu | I >
\end{eqnarray}
In order to distinguish these states, let us look for observables
commutating with $K^\mu_v$. One of these observables is the charge
which is characterized by the phase transformations of the fields
\begin{eqnarray}
Q_v \rightarrow e^{i \alpha} Q_v,  \qquad \bar{Q}_v \rightarrow
e^{- i \alpha} \bar{Q}_v
\end{eqnarray}
The invariance of the Lagrangian under this transformation leads
to a conserved current
\begin{eqnarray}
J^\mu = v^\mu \bar{Q}_v {\xslash v} Q_v
\end{eqnarray}
The space integral of $J^0$ represents the quantum generators of
this transformation
\begin{eqnarray}
Q_c & & = \int d^3 x J^0 = \int d^3 x v^0 : \bar{Q}_v {\xslash v} Q_v : \nonumber \\
& & = \int \frac{d^3 k }{{(2 \pi)}^3 v^0} \sum_s [ b^\dagger_v (
k, s ) b_v ( k, s) - d^\dagger_v (k, s ) d_v (k, s ) ]
\end{eqnarray}
It is not difficult to check the following commutation relations
\begin{eqnarray}
& & [Q_c, b^\dagger_v ( k, s ) ] = b^\dagger_v ( k, s ), \qquad
[ Q_c, b_v ( k, s ) ] = - b_v ( k, s ) \nonumber \\
& & [Q_c, d^\dagger_v ( k, s) ] = - d^\dagger_v ( k, s ), \qquad [
Q_c, d_v ( k, s ) ] = d_v ( k, s )
\end{eqnarray}
and also $[Q_c, K^\mu_v ] = 0$ thus $Q_c$ is time independent and
the theory will describe particles of two types, i.e. particles
and antiparticles. Since the vacuum has zero charge, we then
obtain
\begin{eqnarray}
Q_c |I> = \{^{+ | I >,\  I = 1, 2}_{- | I >,\  I = 3, 4}
\end{eqnarray}

Another observable is spin which is described by the
Pauli-Lubanski operator $W_\sigma$ constructed from the angular
momentum operator $J^{\mu \nu}$. The infinitesimal generator of
Lorentz transformation is given by
\begin{eqnarray}
W_\sigma = - \frac{1}{2} \varepsilon_{\sigma \mu \nu \rho} J^{\mu
\nu} K^{\rho}
\end{eqnarray}
Let $n_\mu = \frac{1}{\sqrt{k^2}} ( | \overrightarrow{k} |, -
\frac{k^0}{ | \overrightarrow{k} | } k_i ) $ be a normalized
space-like four-vector orthogonal to $ k_\mu$,  we can introduce
the helicity operator as
\begin{eqnarray}
J = \frac{W \cdot n}{\sqrt{k^2}}
\end{eqnarray}
Consider now the action of the operator $J$ on the states $|I>$.
The operator $K^\rho$ is replaced by its eigenvalue $k^\rho$.
Choosing the third axis along $\overrightarrow{k}$ and noticing
the following commutation relations of $J^{\mu \nu}$
\begin{eqnarray}
& & [ J^{\mu \nu}_{v^{\prime}}, b^\dagger_v ( k, s ) ] = \delta_{v
v^{\prime}} \int d^3 x \bar{Q}_v ( x )
{\xslash v} \overleftarrow{L}^{\mu \nu} v^0 u ( v, s ) e^{- i k \cdot x }\nonumber \\
& & [ J^{\mu \nu}_{v^{\prime}}, d^\dagger_v ( k, s ) ] = -
\delta_{v v^{\prime}} \int d^3 x \bar{v} ( v, s ) e^{-i k \cdot x}
v^0 L^{\mu \nu} {\xslash v} Q_v ( x )
\end{eqnarray}
and the property $J|0> = 0$, we obtain the following results
\begin{eqnarray}
& & J b^\dagger_v ( k, s ) |0> = \sum_{s^{\prime}} \bar{u} ( v,
s^{\prime} )
\frac{\sigma^{1 2}}{2} u ( v, s ) b^\dagger_v ( k, s^{\prime} ) |0>\nonumber \\
& & J d^\dagger_v ( k , s ) |0> = - \sum_{s^{\prime}} \bar{v} ( v,
s ) \frac{\sigma^{1 2}}{2} v ( v, s^{\prime} ) d^\dagger_v ( k,
s^{\prime} ) | 0>
\end{eqnarray}
Using the relation
\begin{eqnarray}
\bar{u} ( v, s^{\prime} ) \frac{\sigma^{\mu \nu}}{2} u ( v, s ) =
\bar{v} ( v, s^\prime ) \frac{\sigma^{\mu \nu }}{2} v ( v, s ) =
\frac{1}{2} \varepsilon^{\mu \nu \rho \sigma} v_\rho s^i_\sigma
\tau^i_{s s^\prime}
\end{eqnarray}
where $\tau^i$ are the Pauli-matrix and $s^i_\sigma$ are three
normalized space-like four-vector orthogonal to $v_\mu$. Choosing
$v_\mu = ( 1, 0, 0, 0 )$, we have
\begin{eqnarray}
J | I > = \{ ^{+ \frac{1}{2} | I >, \ I = 1, 4}_{- \frac{1}{2}
|I>, \ I = 2, 3 }
\end{eqnarray}

With the results given in eqs.(3.29) and (3.34), we complete the
characterization of states: $|1>$ has charge $+1$ and helicity
$+\frac{1}{2}$, $|2>$ charge $+1$ and helicity $- \frac{1}{2}$, $
|3>$ charge $-1$ and helicity $+\frac{1}{2}$ and $|4>$ charge $-1$
and helicity $- \frac{1}{2}$. For convenience we may denote the
states as
\begin{eqnarray}
& & |I>=|e,v, k, s >, \qquad ( e = \pm, s = \pm \frac{1}{2}, I =1,2,3,4 )\nonumber \\
& & |1> = | +, v, k, \frac{1}{2} >,  \qquad |2> = | +, v, k, - \frac{1}{2} > \nonumber \\
& & |3> = | -, v, k, \frac{1}{2} >, \qquad |4> = | -, v, k, -
\frac{1}{2} >
\end{eqnarray}

It becomes clear for the structure of the Hilbert space of the
HQEFT. The full Hilbert space is the Fock space. Consider now
multi-particle states. Denote the creation operators by the
collective symbol $a^\dagger_I$, with $I = 1,2,3,4$. A basis of
Fock space is generated by the states
\begin{eqnarray}
a^\dagger_{I_1} ( 1 ) ... a^\dagger_{I_n}(n) |0>
\end{eqnarray}
From the anticommutation properties of the creation operators,
these states will be antisymmetric in the wave function argument
$1, ..., n$. In particular, they will vanish if two of those
coincide.

\subsection{Propagator in HQEFT }

From the commutation rules of operators $b_v, b^\dagger_v, d_v $
and $d^\dagger_v$, the anticommutator of two free fields at
arbitrary separations reads
\begin{eqnarray}
& & \{Q^\alpha_v ( x ), \bar{Q}^\beta_{v^\prime} ( x^\prime ) \} \\
& & = \delta_{v v^\prime} \int \frac{d^3 k}{{(2 \pi)}^3 v^0 }
\sum_s [ e^{-i k \cdot (x -x^\prime) } u^\alpha (v, s )
\bar{u}^\beta ( v, s ) + e^{i k (x - x^\prime) } v^\alpha (v, s )
\bar{v}^\beta (v, s ) ]\nonumber
\end{eqnarray}
where $\alpha$ and $\beta$ are Dirac indices. Using the
completeness in eq.(3.23) and writing the phase space measure as
\begin{eqnarray}
\frac{d^3 k}{{(2 \pi)}^3 v^0} = \frac{d^4 k }{{(2 \pi)}^4 } 2 \pi
k \cdot v \delta ( {(k \cdot v)}^2 ) \varepsilon ( k \cdot v )
\end{eqnarray}
with $\varepsilon ( u ) = \frac{u}{| u | }$, we then obtain
\begin{eqnarray}
\{Q^\alpha_v ( x ), \bar{Q}^\beta_{v^\prime} ( x^\prime ) \} =
\delta_{v v^\prime} {\xslash v}_{\alpha \beta} i v \cdot
\partial_x [ i \Delta ( x - x^\prime ) ]
\end{eqnarray}
where
\[ \Delta ( x -x^\prime ) = -i \int \frac{d^4 k }{{(2 \pi)}^4 } \delta ( {(k \cdot v)}^2 )
\varepsilon ( k \cdot v ) \frac{1}{2} [ e^{- i k ( x - x^\prime )
}-e^{i k ( x-x^\prime ) } ] \] If $x^0 = {x^\prime}^0 $, it is
reduced to be
\begin{eqnarray}
\{Q^\alpha_v ( x ), \bar{Q}^\beta_{v^\prime} ( x^\prime ) \}
|_{x^0 = {x^\prime}^0 } =\delta_{v v^\prime } {\xslash v}_{\alpha
\beta} v^{-1}_0 \delta^3 ( \overrightarrow{x}-
\overrightarrow{x}^\prime )
\end{eqnarray}
which agrees with eq.(3.25).

The propagator is defined
\begin{eqnarray}
i S_{v v^\prime} ( x - x^\prime ) = < 0 | {\cal T} Q_v ( x )
\bar{Q}_{v^\prime} ( x^\prime ) | 0 >
\end{eqnarray}
From the definition of the time-ordered product and the
anticommutator relation of the effective fields, it is found that
the propagator satisfies
\begin{eqnarray}
i {\xslash v } v \cdot \partial_x S_{v v^\prime} ( x - x^\prime )
= \delta_{v v^\prime } \delta^4 ( x -x^\prime )
\end{eqnarray}
Its solution is easily read
\begin{eqnarray}
& & < 0 | {\cal T} Q_v ( x ) \bar{Q}_{v^\prime} ( x^\prime ) | 0 >
= iS_{v v^\prime} ( x - x^\prime ) = \delta_{v v^\prime} \int
\frac{d^4 k}{{(2 \pi)}^4} e^{- i k ( x - x^\prime ) }
\frac{i}{{\xslash v } v \cdot k + i \varepsilon }  \nonumber \\
& & \equiv \delta_{v v^\prime} \int \frac{d^4 k}{{(2 \pi)}^4}
\frac{i}{{\xslash v } v \cdot k + i \varepsilon } \left( P_+ e^{-
i{\xslash v } k ( x - x^\prime ) } + P_- e^{ i{\xslash v } k ( x -
x^\prime ) } \right)
\end{eqnarray}
with $P_{\pm} = (1 \pm {\xslash v })/2$ the project operators.

\subsection{Discrete Symmetries in HQEFT }

The discrete symmetries are the symmetries under the parity
transformation, charge conjugation and time reversal, which are
well defined in full QCD. Here we shall extend to the HQEFT of
LCQCD.
\\
\\
i). Parity
\\
\\
Let ${\cal P}$ the unitary operator of parity transformation, from
the field point of view $ {\cal P }$ acting on the field satisfies
in the full theory
\begin{eqnarray}
{\cal P} Q ( x ) {\cal P }^\dagger = \eta_p \gamma^0 Q (\tilde{x}
), \qquad \tilde{x}^\mu = x_\mu, \quad |\eta_p| = 1
\end{eqnarray}
It is expected that ${\cal P}$ acts on the field $Q_v$ to give an
analogous form. What we need to show is how the vector $v_\mu$
transforms under the ${\cal P }$. For that we may notice the fact
that
\begin{eqnarray}
\gamma^0 u ( \tilde{v}, s ) = u ( v, s ), \qquad \gamma^0 v (
\tilde{v} , s ) = - v ( v, s )
\end{eqnarray}
which indicates
\begin{eqnarray}
{\cal P } Q_v ( x ) {\cal P }^\dagger = \eta_p \gamma^0
Q_{\tilde{v}} ( \tilde{x} )
\end{eqnarray}
hence
\begin{eqnarray}
{\cal P } b_v ( k, s ) {\cal P }^\dagger = \eta_p b_{\tilde{v}} (
\tilde{k} , s ), \qquad {\cal P } d_v ( k, s ) {\cal P}^\dagger =
- \eta^\ast_p d_{\tilde{v}} ( \tilde{k} , s )
\end{eqnarray}
Since the $ \gamma^0 Q_{\tilde{v}} ( \tilde{x} ) $ satisfies the
parity transformed equation of motion
\begin{eqnarray}
i {\xslash v} v \cdot \partial [ \gamma^0 Q_{\tilde{v}} (
\tilde{x} ) ] = \gamma^0 i \not{\tilde{v}} \tilde{v} \cdot
\tilde{\partial} [Q_{\tilde{v}} ( \tilde{x} ) ]= 0
\end{eqnarray}
It is then expected that ${\cal P }$ commutes with Hamiltonian
$H$.
\\
\\
ii). Charge Conjugation
\\
\\
In the full theory the charge conjugate operator $ {\cal C}$
acting on the field has the condition
\begin{eqnarray}
{\cal C} Q ( x ) {\cal C}^\dagger = \eta_c C \bar{Q}^T ( x ),
\qquad {\cal C} \bar{Q} ( x ) {\cal C}^\dagger = \eta^\ast_c Q^T (
x ) C
\end{eqnarray}
Where $T$ denotes the transposition acting on the Dirac indices.
$C$ is the combination of $\gamma$ matrices. It is straightforward
to check that this operation is also valid for field $Q_v$, when
\begin{eqnarray}
{\cal C} b_v ( k, s ) {\cal C} = \eta_c d_v ( k, s ), \qquad {\cal
C} d_v ( k, s ) {\cal C} = - \eta^\ast_c b_v ( k, s )
\end{eqnarray}
as $C$ is known satisfying
\begin{eqnarray}
{\cal C} \bar{v}^T( v, s ) = u ( v, s ), \qquad {\cal C} \bar{u}^T
( v , s ) = v ( v, s )
\end{eqnarray}
One can show that $C\bar{Q}^T_v ( x ) $ also satisfies the charge
conjugated equation of motion
\begin{eqnarray}
i {\xslash v } v \cdot \partial [ C \bar{Q}^T_v ( x ) ] = {[ (
\bar{Q}_v i {\xslash v } v \cdot \overleftarrow{\partial} ) C ]}^T
= 0
\end{eqnarray}
\\
iii). Time Reversal
\\
\\
Provided that the field $Q_v ( x )$ satisfies an analogous
transformation as $Q(x )$ does under the time reversal with the
antiunitary operator ${\cal T}$, the transformation of the
four-velocity vector should be clear in this case, since
classically we already know the meaning of time-reversal
invariance is that by reversing the velocities ( space component )
in what used to be the final configuration, a system retraces its
way back to some original configuration, if the fundamental
dynamics has such invariance.

We then expect that ${\cal T}$ acting on $Q_v ( x )$ satisfies
\begin{eqnarray}
{\cal T} Q_v ( x ) {\cal T}^\dagger = \eta_t T Q_{\tilde{v}} ( -
\tilde{x } )
\end{eqnarray}
which requires
\begin{eqnarray}
{\cal T} b_v ( k, s ) {\cal T}^\dagger = \eta_t b_{\tilde{v}} (
\tilde{k}, s ), \qquad {\cal T} d_v ( k, s ) {\cal T}^\dagger = -
\eta^\ast_t d_{\tilde{v}} ( \tilde{k}, s )
\end{eqnarray}
and
\begin{eqnarray}
T u ( v, s ) = u^\ast ( \tilde{v}, s ), \qquad T v ( v, s ) =
v^\ast ( \tilde{v}, s )
\end{eqnarray}
with $T=-i \gamma_5 C$
\\
\\
iv). {${\cal C P T }$ Operation
\\
\\
Let us denote $\Theta = {\cal C P T }$ as the combined
anti-unitary operator which satisfies when acting on field $Q_v (
x )$
\begin{eqnarray}
\Theta Q_v ( x ) \Theta^\dagger = i \gamma_0 \gamma_5 \bar{Q}^T_v
( - x ), \qquad \Theta \bar{Q}_v ( x ) \Theta^\dagger = - Q^T_v (
- x ) i \gamma_0 \gamma_5
\end{eqnarray}

Finally we would like to show that the Lagrangian ${\cal L}_{Q,v}
( x ) $ transforms under $\Theta$ as
\begin{eqnarray}
\Theta {\cal L}_{Q,v} ( x ) \Theta^\dagger = {\cal L}_{Q,v} ( - x
)
\end{eqnarray}
For the each discrete transformation we have
\begin{eqnarray}
 & & {\cal T} {\cal L}_{Q,v} ( x ) {\cal T}^\dagger = {\cal L}_{Q,\tilde{v}} ( - x ) \nonumber \\
 & & {\cal P} {\cal L}_{Q,v} ( x ) {\cal P}^\dagger = {\cal L}_{Q,\tilde{v}} ( \tilde{x} ) \nonumber \\
 & & {\cal C} {\cal L}_{Q,v} ( x ) {\cal C}^\dagger = {\cal L}_{Q,v} ( x )
\end{eqnarray}
Noticing the velocity Lagrangian super-equivalence, we  conclude
that the effective theory is invariance under the parity
transformation, charge conjugation and time reversal.

\section{Basic Framework of HQEFT}

\subsection{Feynman Rules in HQEFT }

The effective Lagrangian describing the interacting system of the
gluon field reads
\begin{eqnarray}
{\cal L }^{(0)}_{Q,v} = \bar{Q}_v i {\xslash v } v \cdot D Q_v
\equiv {\cal L}^{(00)}_{Q,v} + {\cal  L}^{(I)}_{Q,v}
\end{eqnarray}
where
\[{\cal L}^{(00)}_{Q,v} = \bar{Q}_v i {\xslash v} v \cdot \partial Q_v,
\qquad {\cal L}^{(I)}_{Q,v} = \bar{Q}_v i {\xslash v} v^\mu ( -i g
A^a_\mu T^a ) Q_v\]

In the interaction representation and using perturbative theory
technology, it is not difficult to obtain the Feynman rules of the
HQEFT with analogous procedure as in the full QCD theory. The
propagator of heavy quarks is
\begin{eqnarray}
 \frac{i}{\not{v} v \cdot k }
\end{eqnarray}
 and the gluon-heavy quark vertex
\begin{eqnarray}
i g {\xslash v} v_\mu T^a {( 2 \pi )}^4 \delta^4 ( k - k^\prime -
q )
\end{eqnarray}
 The gluon propagator and interactions among them are the same as in the full QCD theory.

\subsection{Effective Lagrangian of HQEFT with Finite Mass corrections}

We now turn to the finite mass case in which the total Lagrangian
can be written as
\begin{eqnarray}
{\cal L}_{HQEFT} = {\cal L}^{(0)}_{Q,v} +  {\cal
L}^{(1/m_Q)}_{Q,v} = {\cal L}^{(00)}_{Q,v} + {\cal L}^{(I)}_{Q,v}
+ {\cal L}^{(1/m_Q)}_{Q,v}
\end{eqnarray}
where ${\cal L}^{(00)}_{Q,v}$ and ${\cal L}^{(I)}_{Q,v}$ are given
in the previous subsection. ${\cal L}^{(1/m_Q)}_{Q,v}$ represents
the finite mass corrections and has the form in terms of the
expansion of inverse powers of heavy quark mass $m_Q$
\begin{eqnarray}
{\cal L}^{(1/m_Q)}_{Q,v} &=&
        \frac{1}{2\MQ} \bar{Q}_v i\DSC \Big(1-\frac{i\DSP}{2\MQ}
\Big)^{-1} i\DSC  Q_v \nonumber \\
          & + &     \frac{1}{2\MQ} \bar{Q}_v (i\CDPNL)  e^{2iv\hspace{-0.15cm}\slash m_Q v\cdot x}
        \Big(1-\frac{i\DSP}{2\MQ}\Big)^{-1} (i\DSCX ) Q_v \\
        & \equiv &   \frac{1}{2\MQ} \bar{Q}_v i\DSC \Big(1-\frac{i\DSP}{2\MQ}
\Big)^{-1} i\DSC  Q_v  \nonumber \\
& + & \frac{1}{2\MQ} \bar{Q}_v (-i\DSCXL)
\Big(1-\frac{-i\oDSP}{2\MQ}
    \Big)^{-1} e^{-2iv\hspace{-0.15cm}\slash m_Q v\cdot x}(i\CDPN) Q_v ,
\end{eqnarray}
where
\begin{eqnarray}
& & i\CDPN=i\DSP+\frac{1}{2\MQ}i\DSC \Big(1-\frac{i\DSP}{2\MQ}
\Big)^{-1} i\DSC
   \nonumber \\
 & & i\CDPNL = -i\oDSP +
 \frac{1}{2\MQ}   (-i\DSCXL)\Big(1-\frac{-i\oDSP}{2\MQ}\Big)^{-1} (-i\DSCXL) .
\end{eqnarray}
Here the finite mass correction term ${\cal L}^{(1/m_Q)}_{Q,v}$
has been expressed by two identities eq.(4.5) and eq.(4.6) for
convenience of use. Their forms are mainly different in the
ordering of operators. The first expression eq.(4.5) is used when
the effective heavy quark field $\bar{Q}_v$ becomes a virtual or
internal one, and the second expression eq.(4.6) is adopted when
the effective heavy quark field $Q_v$ is a virtual or internal
one. This is because the virtual or internal effective heavy quark
will pick up an additional virtual momentum of $ 2 m_Q v $ due to
the factor $e^{\mp 2i v\hspace{-0.15cm}\slash m_Q v\cdot x}$ which
arises from the momentum shift for effective heavy quark and
antiquark fields $\hat{Q}_v^{(\pm)} \to e^{-i
v\hspace{-0.15cm}\slash m_Q v\cdot x} Q_v^{(\pm)} =  e^{\mp i m_Q
v\cdot x} Q_v^{(\pm)}$. While if one uses the effective Lagrangian
given in eq.(2.26), there appears no such an extra momentum
factor.

The $ {\cal L}^{(1/m_Q)}_v $ term should be small in the sense of
effective theory and will be therefore treated as a perturbative
term in the perturbative theory. We would like to point out that
to compute the corrections of $\frac{1}{m_Q}$ to Green functions
and therefore to physical observables one has to add contributions
arising from the mixing terms of the effective quark and antiquark
fields in the effective Lagrangian. In this sense, HQEFT contains
no non-local interactions.

\subsection{Trivialization of Gluon Couplings and Decouple Theorem }

The gluon couplings in the HQEFT of LCQCD can be trivialized in a
similar way as an elimination of the mass term. Such a
trivialization has been shown in QED by Bloch and Nordsieck to
hold in the limit of going on shell. The trivialization in QCD can
be arrived by the change of variable (Wilson-line
transformation)\cite{mainz}
\begin{eqnarray}
& & Q_v = {\cal P} e^{i g \int ^{v \cdot x}_{- \infty} d\tau  v
\cdot A^a T^a} Q^0_v
\equiv W ( x , v ) Q^0_v \nonumber \\
& & \bar{Q}_v = \bar{Q}^0_v {\cal P} e^{- i g \int ^{v \cdot x}_{-
\infty} d\tau  v \cdot A^a T^a } \equiv \bar{Q}^0_v W^{-1}( x, v )
\end{eqnarray}
where ${\cal P}$ denotes path ordering with $x^{\mu} = v^{\mu}
\tau$. Notice that
\begin{eqnarray}
& & v \cdot D Q_v = {\cal P } e^{ i g \int^{v \cdot x }_{- \infty}
v \cdot A^a T^a } v \cdot \partial Q^0_v \nonumber \\
& & ( {\xslash D} - {\xslash v} v \cdot D ) Q_v = {\cal P} e^{ i g
\int^{v \cdot x }_{- \infty} v \cdot A^a T^a } ( {\xslash D} -
{\xslash v} v \cdot D ) Q^0_v
\end{eqnarray}
In terms of the new effective fields, the HQEFT Lagrangian becomes
\begin{eqnarray}
{\cal L}_{Q,v} = \bar{Q}^0_v i {\xslash v} v \cdot \partial Q^0_v
+ {\cal L}^{(1/m_Q)}_{Q,v}
\end{eqnarray}
with
\begin{eqnarray}
    \label{lzz0}
{\cal L}^{(1/m_Q)}_{Q,v} &=&
        \frac{1}{2\MQ} \bar{Q}^0_v i\DSC \Big(1-\frac{i\DSp}{2\MQ}
\Big)^{-1} i\DSC  Q^0_v \nonumber \\
          & + &     \frac{1}{2\MQ} \bar{Q}^0_v (i\CDPNL)  e^{2iv\hspace{-0.15cm}\slash m_Q v\cdot x}
        \Big(1-\frac{i\DSp}{2\MQ}\Big)^{-1} (i\DSCX ) Q^0_v \\
 & \equiv &  \frac{1}{2\MQ} \bar{Q}^0_v i\DSC
\Big(1-\frac{i\DSp}{2\MQ}
\Big)^{-1} i\DSC  Q^0_v  \nonumber \\
& + & \frac{1}{2\MQ} \bar{Q}^0_v (-i\DSCXL)
\Big(1-\frac{-i\oDSp}{2\MQ}
    \Big)^{-1} e^{-2iv\hspace{-0.15cm}\slash m_Q v\cdot x}(i\CDPN) Q^0_v ,
\end{eqnarray}
where
\begin{eqnarray}
& & i\CDPN=i\DSp+\frac{1}{2\MQ}i\DSC \Big(1-\frac{i\DSp}{2\MQ}
\Big)^{-1} i\DSC
   \nonumber \\
 & & i\CDPNL = -i\oDSp +
 \frac{1}{2\MQ}   (-i\DSCXL)\Big(1-\frac{-i\oDSp}{2\MQ}\Big)^{-1} (-i\DSCXL) .
\end{eqnarray}

It becomes manifest that in the infinite mass limit the new
effective field $Q^0_v $ behaves like a free field, which implies
that the heavy quarks decouple from the theory in the infinity
mass limit case. In another word the contributions of the heavy
quarks to the process are suppressed by orders in $1/m_Q$. This
naturally leads to the decouple theorem of heavy quarks in the
strong QCD interactions. This statement holds for both
perturbative and non-perturbative cases.

If there are $N_f$ flavors of heavy quarks in a theory,
transformations indicated in eq.(4.8) with appropriate velocity
dependent paths will lead the effective Lagrangian to a similar
expression except labelled by their own velocity. Note that if
there is only one heavy flavor in a theory, the gluon field can be
simply eliminated by fixing the gauge to be $v \cdot A =0 $. In
the case of $N_f$ flavors of heavy quarks, it can be realized only
in the case that all quarks move at the same velocity. Otherwise
it is not possible to choose a gauge to eliminate such gluon
couplings for all sectors of flavors.

\subsection{Renormalization in HQEFT and Wilson Loops}

Let us now discuss the renormalizability of the HQEFT. To the
leading order in $1/m_Q$, the HQEFT in terms of the effective
field $Q_v$ is already power counting renormalizable. Nevertheless
its renormalizability will be more manifest in terms of the
effective field $Q^0_v$ and Wilson lines. This is because the
Wilson lines have been proved to be renormalizable and the
effective field $Q^0_v$ is a free one that suffer no dressing to
all order of QCD. Note that when including the terms in $1/m_Q$,
i.e. ${\cal L}^{(1/m_Q)}_v$, which is not renormalizable from the
power counting, it is difficult to make precise evaluations of
$1/m_Q$ effects. However, ${\cal L}^{(1/m_Q)}_v $ is suppressed by
the powers of $1/m_Q$ in the sense of effective field theory, it
will be treated as a simple insertion in Green functions in the
perturbative expansion.

To illustrate the renormalization in the HQEFT of LCQCD, we
consider, as a simple example, the renormalization of the operator
\begin{eqnarray}
J^{(0)}_{Q,v} = \bar{Q}^\prime_{v^\prime} ( x ) \Gamma Q_v ( x )
\end{eqnarray}
In terms of the new variable $ Q^0_v$, we have
\begin{eqnarray}
J^{(0)}_{Q,v} = & & \bar{Q}^{\prime 0}_{v^\prime} ( x ) {\cal P}
e^{ - i g \int^{v^\prime \cdot x}_{- \infty} v^\prime \cdot A^a
T^a } \Gamma \
{\cal P } e^{ - i g \int^{v \cdot x }_{- \infty} v \cdot A^a T^a } Q^0_v \nonumber \\
\equiv & & \bar{Q}^{\prime 0}_{v^\prime} W^{-1} ( x, v^\prime ) W
( x, v ) \Gamma Q^0_v
\end{eqnarray}

As shown from above subsection that to the leading order in
$1/m_Q$ the effective field $Q^0_v ( x )$ is with respect to the
free field and therefore suffers no renormalization to all orders
in the coupling. The renormalization of the operator
$J^{eff}_\Gamma$ is then reduced to the renormalization of the
Wilson loops, i.e.
\begin{eqnarray}
\frac{1}{N_c} Tr < 0 | W^{-1} ( x, v^\prime ) W ( x, v ) | 0> =
\frac{1}{N_c } <0| {\cal P} e^{i g \oint_{C_\delta} d x^\mu A_\mu
} | 0> \equiv W ( C_\delta )
\end{eqnarray}
where $ C_\delta$ is the loop with, at the point $v^\prime \cdot x
= v \cdot x$, a cusp characterized by angle $\delta$. In Minkowski
space the angle $\delta$ is given by
\begin{eqnarray}
\cosh \delta = v \cdot v^\prime
\end{eqnarray}

It was shown that a Wilson loop is multiplicatively renormalizable
in the case of having a finite number of self-interaction points
and cusps corresponding to angles $\{ \delta_i\}$. The
renormalization properties of the Wilson loops containing cusp
singularities have been studied in ref.\cite{AP,KR}. The two-loop
contribution to the cusp anomalous dimension $\Gamma_{cusp}(\delta
, g )$ was calculated in \cite{KR}. It was found that the
renormalized contour average $W_R$ of Wilson loop in the presence
of a cusp can be constructed by applying the ordinary ${\cal
R}$-operation.
\begin{eqnarray}
A^\mu \rightarrow A^\mu_R = Z^{-\frac{1}{2}}_3 A^\mu, \qquad g
\rightarrow g_R = Z^{-1}_1 Z^{\frac{3}{2}}_3
\mu^{\frac{\varepsilon}{2}}g
\end{eqnarray}
with incorporating the subtraction procedure ${\cal K}_\delta$
\begin{eqnarray}
W_R ( C_\delta; g_R, \mu, \bar{C}_\delta ) = \lim_{\varepsilon
\rightarrow 0} {\cal K}_\delta \tilde{W} ( C_\delta; g_R, \mu,
\varepsilon ) = \lim_{\varepsilon \rightarrow 0} {\cal K}_\delta
{\cal R} W ( C_\delta; g, \varepsilon )
\end{eqnarray}
where $\bar{C}_\delta$ denotes a generalized subtraction point of
the ${\cal K}_\delta $ procedure. It is also proved that the
renormalized contour average $ W_R ( C_\delta; g_R, \mu,
\bar{C}_\delta )$ satisfies the exponential theorem
\begin{eqnarray}
W_R ( C_\delta; g_R,\mu, \bar{C}_\delta ) = e^{W^{2 P I}_R (
C_\delta; g_R, \mu, \bar{C}_\delta ) }
\end{eqnarray}
where $W^{2 P I }_R$ is the two-particle irreducible contour
averages, and satisfies the renormalization group equation
\begin{eqnarray}
[ \mu \frac{\partial}{\partial \mu} + \beta ( g_R)
\frac{\partial}{\partial g_R} ] W_R ( C_\delta; g_R, \mu,
\bar{C}_\delta ) = - \Gamma_{cusp} ( \delta, g_R )
\end{eqnarray}
$\Gamma_{cusp} ( \delta, g_R ) $ is the anomalous dimension
defined by
\begin{eqnarray}
\Gamma_{cusp} (\delta, g_R ) = - \lim_{\varepsilon \rightarrow 0}
\frac{d}{d ( \ln \mu)} \tilde{W}^{2 P I } ( C_\delta; g_R, \mu,
\varepsilon )
\end{eqnarray}
The general solution of the renormalization group equation can be
written
\begin{eqnarray}
W_R ( C_\delta; g_R, \mu, \bar{C}_\delta ) = W_R ( C_\delta; g_R,
\bar{\mu}, \bar{C}_\delta ) e^{- \int^{g_R (\mu ) }_{g_R (
\bar{\mu} )} d g \frac{\Gamma_{cusp} ( \delta, g ) }{ \beta(g) } }
\end{eqnarray}
Note that the cusp anomalous dimension $\Gamma_{cusp} ( \delta, g
)$ depends only on the cusp angle $\delta$. One- and two-loop
contributions to the cusp anomalous dimension have been calculated
explicitly in ref.\cite{KR}. To be manifest, let us quote , for
example , the one-loop result
\begin{eqnarray}
\Gamma^{one-loop}_{cusp} ( \delta, g_R) = \frac{\alpha_s}{\pi} C_F
( \delta \coth \delta -1 )
\end{eqnarray}
which was also reproduced from directly calculating the one-loop
QCD corrections of heavy quark currents\cite{SW}.

\subsection{Current Operators in HQEFT }

Just like the effective Lagrangian is established from the full
theory, the current operators in the full QCD theory may be
expressed in terms of the new effective fields in HQEFT of LCQCD.
As an example, consider the current operator
\begin{eqnarray}
J_Q= \bar{Q}^\prime ( x ) \Gamma Q ( x )
\end{eqnarray}
which relates to the current operators in the HQEFT of LCQCD via a
sequence of field redefinitions given in the previous sections.
\begin{eqnarray}
J_Q \to & & J_{Q,v} = \bar{Q}^\prime_{v^\prime} ( x )
e^{i\not{v}^\prime m_{Q^\prime} v^\prime \cdot x }
 \Gamma e^{-i \not{v} m_Q v \cdot x } Q_v ( x )\nonumber \\
& & + \bar{Q}^\prime_{v^\prime} ( x ) e^{i \not{v}^\prime
m_{Q^\prime} v^\prime \cdot x }
\Gamma e^{i \not{v} m_Q v \cdot x } W_v ( \frac{1}{m_Q} ) Q_v ( x )\nonumber \\
& & +\bar{Q}^\prime_{v^\prime} ( x ) \overleftarrow{W}_{v^\prime}
( \frac{1}{m_{Q^\prime}} )
e^{-i \not{v}^\prime m_{Q^\prime} v^\prime \cdot x }
\Gamma e^{-i \not{v} m_Q v \cdot x } Q_v (x )\nonumber \\
& & + \bar{Q}^\prime_{v^\prime} ( x ) \overleftarrow{W}_{v^\prime}
( \frac{1}{m_{Q^\prime}} )
 e^{-i \not{v}^\prime m_{Q^\prime} v^\prime \cdot x }
 \Gamma e^{i \not{v} m_Q v \cdot x } W_v ( \frac{1}{m_Q} ) Q_v ( x )
\end{eqnarray}
Noticing the following identity
\begin{eqnarray}
e^{\epsilon^\prime i \not{v}^\prime m_{Q^\prime} v^\prime \cdot x
} \Gamma e^{-\epsilon i \not{v} m_Q v \cdot x} \equiv
\sum_{\sigma^\prime = \pm} \sum_{\sigma = \pm } \frac{1+
\epsilon'\sigma^\prime {\xslash v^\prime} }{2} \Gamma \frac{1 +
\epsilon \sigma {\xslash v }}{2} e^{\epsilon' \sigma^\prime i
m_{Q^\prime} v^\prime \cdot x - \epsilon \sigma i m_Q v \cdot x}
\end{eqnarray}
with $\epsilon'=\pm, \ \epsilon=\pm$.  Then eq.(4.26) can be
rewritten as
\begin{eqnarray}
J_{Q,v} = & & \sum_{\sigma^\prime = \pm} \sum_{\sigma = \pm}
e^{\sigma^\prime i m_{Q^\prime} v^\prime \cdot x - \sigma i m_Q v
\cdot x } [ \bar{Q}^\prime_{v^\prime} ( x ) \frac{1+ \sigma^\prime
{\xslash v^\prime} }{2}
\Gamma \frac{1 + \sigma {\xslash v }}{2} Q_v ( x ) \nonumber \\
& & + \bar{Q}^\prime_{v^\prime} ( x ) \frac{1+ \sigma^\prime
{\xslash v^\prime} }{2}
\Gamma \frac{1 + \sigma {\xslash v }}{2} W_v ( \frac{1}{m_Q} ) Q_v ( x )\nonumber \\
& & +\bar{Q}^\prime_{v^\prime} ( x ) \overleftarrow{W}_{v^\prime}
( \frac{1}{m_{Q^\prime}} ) \frac{1+ \sigma^\prime {\xslash
v^\prime} }{2}
\Gamma \frac{1 + \sigma {\xslash v }}{2} Q_v ( x ) \nonumber \\
& & +\bar{Q}^\prime_{v^\prime} ( x ) \overleftarrow{W}_{v^\prime}
( \frac{1}{m_{Q^\prime}} ) \frac{1+ \sigma^\prime {\xslash
v^\prime} }{2} \Gamma \frac{1 + \sigma {\xslash v }}{2} W_v (
\frac{1}{m_Q} ) Q_v ( x ) ] \nonumber \\
\equiv & &  J_{Q,v}^{(0)} + J_{Q,v}^{(1/m_Q)}
\end{eqnarray}
where the summation contains four cases which correspond to the
transitions between quarks and between antiquarks  as well as the
creation and annihilation of the quark and antiquark pairs.

Beyond the tree level, the above expression has to be replaced by
a more general sum over operators. Note that in computing the
coefficient functions of operators to the order in $1/m_Q$ one
should include the graphs with an insertion of the terms in the
same order of $1/m_Q$ from the effective Lagrangian. In
particular, we would like to stress that in the computation of the
coefficient functions of the terms of order in $1/m_Q$, there is
also a contribution from the terms of the quark-antiquark coupling
terms in the effective Lagrangian. For instance, integrating over
the antiquark field, the resulting effective current for quark
field receive contributions starting from the order of $1/m_Q$. In
general, the effective current can be written as follows after
integrating over the antiquark field
\begin{eqnarray}
J_{eff}^{(++)} = J_{eff}^{(0)} + J_{eff}^{(1/m_Q)}
\end{eqnarray}
with
\begin{eqnarray}
 & &  J_{eff}^{(0)} = e^{i(m_{Q^{\prime}} v^{\prime} -m_Q v)\cdot x} \QVBP \Gamma
 \QV  \\
 & &  J_{eff}^{(1/m_Q)} = e^{i(m_{Q^{\prime}} v^{\prime}
        -m_Q v)\cdot x}
       \{ \frac{1}{2\MQ}\QVBP\Gamma\frac{1}{i\DSP}(i\DSC)^2 \QV  \nonumber\\
        &&+\frac{1}{2\MQP}\QVBP (-i\stackrel{\hspace{-0.1cm}\leftarrow}{\DSC})^2\frac{1}
        {-i{\oDSP}}\Gamma\QV+\frac{1}{4\MQ^2}\QVBP
        \Gamma\frac{1}{i\DSP}
        i\DSC (i\DSP)  \nonumber\\
        &&\times i\DSC \QV
        +\frac{1}{4\MQP^2}\QVBP (-i\stackrel{\hspace{-0.1cm}\leftarrow}{\DSC})
        (-i{\oDSP})(-i\stackrel{\hspace{-0.1cm}\leftarrow}{\DSC})
        \frac{1}{-i{\oDSP}}\Gamma \QV  \\
        &&+\frac{1}{4\MQP \MQ}\QVBP (-i\stackrel{\hspace{-0.1cm}\leftarrow}{\DSC})^2
        \times \frac{1}{-i{\oDSP}}\Gamma
        \frac{1}{i\DSP}(i\DSC)^2 \QV
        +O(\frac{1}{m_{Q^{(\prime)}}^3}) \}  \nonumber
\end{eqnarray}
where we only keep to the order of $1/m_Q^2$. We have also used
the operator condition $ (i\DSC)^2/2m_Q  \ll i\DSP $ in the heavy
quark expansion, which holds for the typical case with $ \langle
i\DP \rangle \sim \langle i\DC \rangle \sim \bar{\Lambda} \ll m_Q$
in the heavy-light hadron system.

We would like to address that the above forms of effective current
are quite different from those without considering the
contributions of antiquark fields. Especially, the terms
$\frac{1}{2\MQ}\QVBP \Gamma i\DSC \QV$ and
$\frac{1}{2\MQP}\QVBP(-i\stackrel{\hspace{-0.15cm}\leftarrow}{\DSC})\Gamma
\QV$ disappear in HQEFT. This is because they are exactly
cancelled by the additional contributions arising from the
intermediate antiquark fields. As a consequence, the operator
forms in the above effective current $J_{eff}^{(++)}$ and those in
the effective Lagrangian $L^{(++)}_{eff}$ become similar, namely
all the odd powers of the transverse momentum operator $\DSC$ are
absent, only the even powers of $\DSC$ appear in the effective
current $J_{eff}^{(++)}$ and effective Lagrangian
$L^{(++)}_{eff}$. Such an interesting feature in HQEFT becomes
remarkable in evaluating the hadronic matrix elements. For
instance, fewer form factors are involved, and $1/m_Q$ corrections
at zero recoil are automatically absent for both transitions
between heavy pseudoscalar to heavy vector and between heavy
pseudoscalar to heavy pseudoscalar mesons.


\subsection{Spin and Angular Momentum in HQEFT}

To discuss the spin and rotational symmetries in HQEFT of LCQCD,
it is useful to construct the generators by introducing the
following Pauli-Lubanski vector
\begin{eqnarray}
 \label{4.32}
W_\sigma=-\frac{1}{2} \varepsilon_{\sigma \mu \nu \rho} J^{\mu
\nu} v^\rho
\end{eqnarray}
where $J^{\mu \nu}$ is the angular momentum operator, i.e. the
infinitesimal generator of Lorentz transformation.
\begin{eqnarray}
\label{4.33}
 J^{\mu \nu} = \frac{1}{2} \sigma^{\mu \nu}+i (x^\mu
\partial^\nu - x^\nu \partial^\mu)
\end{eqnarray}
Substituting (\ref{4.33}) into (\ref{4.32}), we obtain
\begin{eqnarray}
W_\sigma = S_\sigma + L_\sigma
\end{eqnarray}
with
\begin{eqnarray}
& & S_\sigma = -\frac{1}{4} \varepsilon_{\sigma \mu \nu \rho}
\sigma^{\mu \nu} v^\rho = \frac{i}{2} \gamma_5 \sigma_{\sigma \rho} v^\rho
= - \frac{1}{2} \gamma_5 ( \gamma_\sigma - v_\sigma {\xslash v} ) {\xslash v } \nonumber \\
& & L_\sigma = - \frac{i}{2} \varepsilon_{\sigma \mu \nu \rho} (
x^\mu \partial^\nu - x^\nu \partial^\mu ) v^\rho
\end{eqnarray}
It is not difficult to show that the effective Lagrangian in the
infinite mass limit is invariant under the transformations
\begin{eqnarray}
Q_v \rightarrow e^{i S_\sigma \alpha^\sigma}, \qquad
\bar{Q}_v \rightarrow \bar{Q}_v e^{-i S_\sigma \alpha^\sigma}\\
Q_v \rightarrow e^{i L_\sigma \alpha^\sigma}, \qquad \bar{Q}_v
\rightarrow \bar{Q}_v e^{-i L_\sigma \alpha^\sigma}
\end{eqnarray}
and therefore
\begin{eqnarray}
Q_v \rightarrow e^{i W_\sigma \alpha^\sigma }, \qquad \bar{Q}_v
\rightarrow \bar{Q}_v e^{- i W_\sigma \alpha^\sigma }
\end{eqnarray}

At the rest frame with $v = ( 1, 0, 0, 0 )$, the angular momentum
can be written as
\begin{eqnarray}
W_i = L_i + S_i
\end{eqnarray}
with
\begin{eqnarray}
& & S_i =\frac{1}{4} \varepsilon_{i j k} \sigma^{j k} =
\frac{1}{2} \gamma^5 \gamma^0 \gamma^i = \frac{1}{2}
\left ( \begin{array}{cc}\sigma^i & 0\\ 0 & \sigma^i \end{array} \right ) \nonumber \\
& & L_i = \frac{i}{2} \varepsilon_{i j k} ( x^j \partial^k - x^k
\partial^j)
\end{eqnarray}
where $S_i$ are the usual spin-matrices and $L_i$ are the usual
orbital angular momentum operators. This shows that the spin
$\overrightarrow{S}$ and orbital angular momentum
$\overrightarrow{L}$ are separately conserved, angular momentum
$\overrightarrow{W}$ is therefore also conserved in infinite mass
limit.

\subsection{New Symmetries in HQEFT at infinite mass limit}

In addition to spin-flavor symmetry, we shall show that the
effective Lagrangian in infinite mass limit is also invariant
under the following transformations
\begin{eqnarray}
& & Q_v \rightarrow e^{i ( \alpha_5\gamma_5 + \alpha ) } Q_v,
\qquad
\bar{Q}_v \rightarrow \bar{Q}_v e^{ i ( \alpha_5 \gamma_5 - \alpha ) }\\
& & Q_v \rightarrow e^{i \not{v}( ia_5\gamma_5 + a ) } Q_v, \qquad
\bar{Q}_v \rightarrow \bar{Q}_v e^{ i \not{v}( a_5 \gamma_5 - a) }\\
& & Q_v \rightarrow e^{i ( \gamma_\mu - v_\mu \not{v} ) ( b_5^\mu
\gamma_5 + i b^\mu )} Q_v,
 \qquad \bar{Q}_v \rightarrow \bar{Q}_v e^{- i ( \gamma_\mu - v_\mu \not{v} )
 ( b_5^\mu \gamma_5 - i b^\mu ) }\\
& & Q_v \rightarrow e^{i ( v_\mu \gamma_\nu - v_\nu \gamma_\mu )
( c_5^{\mu \nu} \gamma_5 + i c^{\mu \nu} ) } Q_v\nonumber \\
& & \bar{Q}_v \rightarrow \bar{Q}_v e^{-i ( v_\mu \gamma_\nu - v_\nu \gamma_\mu )
( c_5^{\mu \nu}\gamma_5 - i c^{\mu \nu} ) }\\
& & Q_v \rightarrow e^{i [ \sigma_{\mu \nu} + i ( v_\mu \gamma_\nu
- v_\mu \gamma_\nu ) \not{v} ]
( d_5^{\mu \nu} \gamma_5 + d^{\mu \nu} )} Q_v \nonumber \\
& & \bar{Q}_v \rightarrow \bar{Q}_v e^{i [ \sigma_{\mu\nu} + i (
v_\mu \gamma_\nu - v_\nu \gamma_\mu ) \not{v} ]
 ( d_5^{\mu \nu} \gamma_5 - d^{\mu \nu} )} \\
& & Q_v \rightarrow e^{i \not{v}\not{\varepsilon} ( f _5 \gamma_5
+ f ) } Q_v,
\qquad \bar{Q}_v \rightarrow \bar{Q}_v e^{- i \not{v} \not{\varepsilon} ( f_5 \gamma_5 - f ) } \\
& & Q_v \rightarrow e^{i \gamma_5 \not{\varepsilon} g } Q_v,
\qquad \bar{Q}_v \rightarrow \bar{Q}_v e^{- i \gamma_5
\not{\varepsilon} g }
\end{eqnarray}
where $\varepsilon_{\mu}$ is the polarization vector with
$\varepsilon\cdot v = 0$.

\section{Hadronic Matrix Elements and $1/m_Q$ Expansions in HQEFT }

 As a direct application of HQEFT, we are going to show how the
hadronic matrix element in full QCD theory can systematically be
expanded into a series of matrix elements in terms of $1/{\MQ}$ in
HQEFT.

\subsection{Weak Transition Matrix Elements in HQEFT }

Adopting the conventional relativistic normalization
      \begin{equation}
         <H(p^{\prime})\vert H(p)>=2p^{0}(2\pi)^3 \delta^{3}({\vec{p}}^{\hspace{0.1cm}\prime}
         -\vec{p}),
      \end{equation}
the conservation of the vector current $\bar{Q}\gamma^{\mu}Q$
leads to
      \begin{equation}
      \label{5.2}
          <H(p)\vert \bar{Q}\gamma^{\mu}Q \vert H(p)>=2p^{\mu}=2m_H v^{\mu},
      \end{equation}
where $\vert H>$ denotes a hadron state in QCD and $p^{\mu}=m_H
v^{\mu}$ is the momentum of the heavy hadron $H$. This may be
regarded as an alternative definition of heavy hadron mass.

To exhibit a manifest spin-flavor symmetry in HQEFT at infinite
mass limit, we introduce an effective heavy hadron state $\vert
H_v>$ with the normalization
\begin{equation}
 \label{5.3}
         <H_{v} \vert \bar{Q}_v \gamma^{\mu} Q_v \vert H_v> = 2\bar{\Lambda} v^{\mu}
     \end{equation}
where
\[ \bar{\Lambda} = \lim_{m_{Q}\to \infty} \bar{\Lambda}_H \]
is a heavy flavor-independent binding energy that reflects the
confinement effects of the light degrees of freedom in the heavy
hadron.

The hadronic matrix element in full QCD is then given by the one
in HQEFT via the following formulation
    \begin{equation}
    \label{5.4}
     \frac{1}{\sqrt{m_{H^{\prime}}m_{H}}} <H^{\prime}\vert J_Q \vert H>=
\frac{1}{\sqrt{\bar{\Lambda}_{H^{\prime}} \bar{\Lambda}_H}}
<H^{\prime}_{v^{\prime}}\vert
          J_{Q,v} e^{i\int d^4x {\cal L}_{HQEFT}}\vert H_v>.
    \end{equation}
Here $\bar{\Lambda}_{H}$ and $\bar{\Lambda}_{H'}$ are the binding
energies defined as
    \begin{equation}
       \bar{\Lambda}_{H} \equiv m_{H}-m_{Q} ,\hspace{1.5cm}
       \bar{\Lambda}_{H^{\prime}} \equiv m_{H^{\prime}}-m_{Q^{\prime}}.
    \end{equation}
The flavor-dependent factor
$\frac{1}{\sqrt{\bar{\Lambda}_{H^{\prime}} \bar{\Lambda}_H}}$
appears due to the different normalization of the hadron states
$\vert H>$ in full QCD and $\vert H_v>$ in HQEFT.

In general, the $1/{m_Q}$ corrections to the hadronic matrix
elements in the heavy quark expansion can be classified into three
parts: 1) corrections from purely effective current
$J^{(1/{\MQ})}_{Q,v}$; 2) corrections from purely effective
Lagrangian ${\cal L}^{(1/{\MQ})}_{Q,v}$; and 3) mixed corrections
from both $J^{(1/{\MQ})}_{Q,v}$ and $L^{(1/{\MQ})}_{Q,v}$.

After a detailed evaluation with contracting the effective heavy
quark field, the hadronic matrix element in HQEFT is found to have
the following general form up to order of $1/m_Q^2$
\begin{eqnarray}
\label{matrixA}
   {\cal A} &\equiv & <H^{\prime}_{v^{\prime}}\vert
          J_{Q,v} e^{i\int d^4x {\cal L}_{HQEFT}}\vert H_v>
         \nonumber\\
        &=& <H^{\prime}_{v^{\prime}}\vert \QvBP \Gamma \Qv \vert H_v>
        -\frac{1}{2\MQ}<H^{\prime}_{v^{\prime}}\vert \QvBP O_{1}(\Gamma) \Qv \vert H_v>\nonumber\\
        &&-\frac{1}{2\MQP}<H^{\prime}_{v^{\prime}}\vert \QvBP O_{1}^{\prime}(\Gamma)
        \Qv \vert H_v>
        -\frac{1}{4\MQ^2}<H^{\prime}_{v^{\prime}}\vert \QvBP
        O_{2}(\Gamma)\Qv
        \vert H_v>  \nonumber\\
        &&-\frac{1}{4\MQP^2}<H^{\prime}_{v^{\prime}}\vert \QvBP
        O_{2}^{\prime}(\Gamma)\Qv
        \vert H_v>
        -\frac{1}{4\MQ^2}<H^{\prime}_{v^{\prime}}\vert \QvBP O_{3}(\Gamma)\Qv \vert H_v>
       \nonumber  \\
        &&-\frac{1}{4\MQP^2}<H^{\prime}_{v^{\prime}}\vert \QvBP
        O_{3}^{\prime}(\Gamma)\Qv
        \vert H_v>
        +\frac{1}{4\MQP \MQ}<H^{\prime}_{v^{\prime}}\vert \QvBP O_{4}(\Gamma)\Qv \vert H_v>
      \nonumber \\
         &&+O(\frac{1}{m_{Q^{\prime}}^3}) .
\end{eqnarray}
which explicitly displays the $1/m_Q$ expansion. Where the
operators $O_{i}(\Gamma)$ and  $O'_{i}(\Gamma)$ are defined as
follows
   \begin{eqnarray}
     O_1(\Gamma)& =&\Gamma\frac{1}{i\DSPR}(i\DSC)^2 ,\nonumber\\
     O_1^{\prime}(\Gamma)&=&(-i\stackrel{\hspace{-0.1cm}\leftarrow}{\DSC})^2
\frac{1}{-i {\DSPL}} \Gamma ,   \nonumber \\
     O_2(\Gamma)& =&\Gamma\frac{1}{i\DSPR}(i\DSC)(i\DSP)i\DSC , \nonumber\\
     O_2^{\prime}(\Gamma)& =&(-i\stackrel{\hspace{-0.1cm}\leftarrow}{\DSC})
(-i{\DSPl})
        (-i\stackrel{\hspace{-0.1cm}\leftarrow}{\DSC})
\frac{1}{-i{\DSPL}}
        \Gamma ,\nonumber\\
     O_3(\Gamma) &=&\Gamma\frac{1}{i\DSPR}(i\DSC)^2 \frac{1}{i\DSPR}
        (i\DSC)^2 , \nonumber\\
     O_3^{\prime}(\Gamma)& =&(-i\stackrel{\hspace{-0.1cm}\leftarrow}{\DSC})^2
\frac{1}{-i{\DSPL}}(-i\stackrel{\hspace{-0.1cm}\leftarrow}{\DSC})^2
\frac{1}{-i{\DSPL}} \Gamma ,\nonumber\\
     O_4(\Gamma)& =&(-i\stackrel{\hspace{-0.1cm}\leftarrow}{\DSC})^2
\frac{1}{-i{\DSPL}}\Gamma\frac{1}{i\DSP}(i\DSC)^2.
   \end{eqnarray}
which are all given in the even powers of $\DSC$. Note that the
term $\frac{1}{i\DSPR}$ (or $\frac{1}{i\DSPL}$) arises with
replacing the propagator from contracting effective heavy quark
fields in HQEFT by the corresponding operator.

\subsection{Mass Formula of Hadrons and Transition Matrix Elements
at Zero Recoil in HQEFT}

  Based on the formulation eq.(5.4) and the normalization conditions for hadron states in
  full QCD (eq.(5.2)) and in HQEFT (eq.(5.3)), by setting $v^{\prime}=v$, we obtain
    \begin{eqnarray}
       2m_H v^{\mu}&=&<H\vert Q\gamma^{\mu} Q \vert H>=\frac{m_H}{\bar{\Lambda}_H}
         \{2\bar{\Lambda} v^{\mu}-\frac{1}{\MQ}<H_v\vert \QvB O_{1}(\gamma^{\mu})
         \Qv \vert H_v> \nonumber\\
       &&-\frac{1}{2\MQ^2}<H_v\vert \QvB (O_{2}(\gamma^{\mu})+O_{3}(\gamma^{\mu}))
         \Qv \vert H_v> \nonumber\\
       &&+\frac{1}{4\MQ^2}<H_v\vert \QvB O_{4}(\gamma^{\mu})\Qv \vert H_v>
         +O(\frac{1}{m_{Q}^3})  \} ,
    \end{eqnarray}
  which allows us to extract the binding energy in terms of
  $1/m_Q$ expansion
    \begin{eqnarray}
       \bar{\Lambda}_{H}&=&\bar{\Lambda}-\frac{1}{2\MQ}<H_v\vert \QvB O_{1}(\VS)
        \Qv \vert H_v>
       -\frac{1}{4\MQ^2}<H_v\vert \QvB (O_{2}(\VS)+O_{3}
         (\VS)\Qv \vert H_v> \nonumber\\
      &&+\frac{1}{8\MQ^2}<H_v\vert \QvB O_{4}(\VS)\Qv \vert H_v>
        +O(\frac{1}{\MQ^3}) .
    \end{eqnarray}
 We then arrive at an alternative definition for the heavy hadron mass
    \begin{equation}
      m_H=m_Q +\bar{\Lambda}_H=m_Q +\bar{\Lambda}+ O(1/{m_Q}),
    \end{equation}
  which shows that the mass of a hadron is given by three parts:
  the effective heavy quark mass $m_Q$, the binding energy $\bar{\Lambda}$ due to light
  degrees of freedom and terms suppressed by $1/{\MQ}$.

We now turn to discuss the transition matrix elements. For
illustration, consider first a simple case that both the initial
and final states are pseudoscalar mesons. We may choose a
realistic process, i.e., $B\rightarrow D$ transition matrix
element for vector current. From the formulation eq.(5.4) and the
transition matrix elements in the heavy quark expansion eq.
(\ref{matrixA}), it is easily read
  \newcommand{\CVP}{\bar{c}^{+}_{v^{\prime}}}
  \newcommand{\BV}{b_v^{+}}
   \begin{eqnarray}
   \label{BDTME}
      <D\vert \bar{c} \gamma^{\mu} b \vert B &&> =
      \sqrt{\frac{m_D m_B}{\bar{\Lambda}_D \bar{\Lambda}_B}}
        \{  <D_{v^{\prime}}\vert \CVP \gamma^{\mu} \BV \vert B_v>
        -\frac{1}{2m_b}<D_{v^{\prime}}\vert \CVP O_{1}(\gamma^{\mu}) \BV \vert B_v>\nonumber\\
       && -\frac{1}{2m_c}<D_{v^{\prime}}\vert \CVP O_{1}^{\prime}(\gamma^{\mu})
        \BV \vert B_v> -\frac{1}{4m_b^2}<D_{v^{\prime}}\vert \CVP O_{2}(\gamma^{\mu})\BV
        \vert B_v> \nonumber\\
       &&-\frac{1}{4m_c^2}<D_{v^{\prime}}\vert \CVP O_{2}^{\prime}(\gamma^{\mu})\BV
        \vert B_v>
        -\frac{1}{4m_b^2}<D_{v^{\prime}}\vert \CVP O_{3}(\gamma^{\mu})\BV \vert B_v> \nonumber \\
       && -\frac{1}{4m_c^2}<D_{v^{\prime}}\vert \CVP O_{3}^{\prime}(\gamma^{\mu})\BV
        \vert B_v>
       +\frac{1}{4m_c m_b}<D_{v^{\prime}}\vert \CVP O_{4}(\gamma^{\mu})\BV \vert B_v> \nonumber\\
       && +O(\frac{1}{m_{b(c)}^3})  \} .
   \end{eqnarray}

Applying spin-flavor symmetry, we have the following relations for
operators $O_i$
    \begin{equation}
    \label{BDTMESP}
      <B_v \vert \bar{b}^{+}_{v}O_i b^{+}_v\vert B_v>
        =<D_v \vert \bar{c}^{+}_{v}O_i c^{+}_v
      \vert D_v> \equiv <P_v \vert \QVB O_i \QV \vert P_v>.
    \end{equation}
With the above relations and the binding energy relations
eq.(5.9), it is not difficult to find that the $B\to D$ transition
matrix element is simplified to the following form at zero recoil
$v^{\prime}=v$,
    \begin{eqnarray}
      <D\vert \bar{c}\gamma^{\mu} b\vert B>|_{q^{2}_{max}}
         & & = 2\sqrt{m_B m_D}v^{\mu}\{1+\frac{1}{32\bar{\Lambda}^{2}}
      (\frac{1}{m_b}-\frac{1}{m_c})^2 {<P_v\vert \QVB O_{1}(\VS)\QV \vert P_v>}
             \nonumber\\
      && -\frac{1}{16\bar{\Lambda}}(\frac{1}{m_b}-\frac{1}{m_c})^2
       <P_v\vert \QVB O_{4}(\VS)\QV \vert P_v> +O(\frac{1}{m_{b(c)}^3}) \} ,
    \end{eqnarray}
which explicitly shows that when applying the formulation eq.(5.4)
and the normalizations eq.(5.2) and eq.(5.3), the transition
matrix elements of heavy quark vector current between two
pseudoscalar mesons automatically do not receive corrections of
order $1/{\MQ}$ at zero recoil even without analyzing the concrete
Lorentz structure and evaluating the hadronic matrix elements for
each operator.

\subsection{Trace Formula of Transition Matrix Elements and
Universal Isgur-Wise Function }

Evaluating the transition matrix elements is a hard task. The
transition matrix element for heavy-light hadrons is generally
defined as
\begin{eqnarray}
T_{\Gamma} = <H' (p')| \int_z \bar{Q}^\prime ( z ) \Gamma Q ( z )
e^{i q \cdot z } | H(p) >
\end{eqnarray}
 For illustration, we first consider its leading term in
 HQEFT, which is simply given by
\begin{eqnarray}
T_{\Gamma}^0 & = & <H' _{v'} | \int_z \bar{Q}_{v^\prime}^\prime ( z )
e^{i \not{v^\prime} m_{Q^\prime} v^\prime \cdot z } \Gamma e^{ i \not{v} m_Q v \cdot z }
 Q_v ( z ) e^{ i q \cdot z } | H_v >\nonumber \\
&  = & <H' _{v'}| \int_z \bar{Q}_{v^\prime}^\prime ( z ) \frac{1+{\xslash v^\prime}}{2}
 \Gamma \frac{1+{\xslash v}}{2} Q_v ( z ) e^{i (q - m_Q v
 + m_{Q^\prime} v^\prime ) \cdot z } |H_v >\nonumber \\
& +  & <H' _{v'} | \int_z \bar{Q}_{v^\prime}^\prime ( z )
\frac{1-{\xslash v^\prime}}{2}\Gamma \frac{1-{\xslash v}}{2}
Q_v ( z ) e^{i (q + m_Q v - m_{Q^\prime} v^\prime ) \cdot z } |H_v >\nonumber \\
& + & <H' _{v'} | \int_z \bar{Q}_{v^\prime}^\prime ( z ) \frac{1-{\xslash v^\prime}}{2}
 \Gamma \frac{1+{\xslash v}}{2} Q_v ( z )
 e^{i (q - m_Q v - m_{Q^\prime} v^\prime ) \cdot z } |H_v >\nonumber \\
& + & <H' _{v'} | \int_z \bar{Q}_{v^\prime}^\prime ( z )
\frac{1+{\xslash v^\prime}}{2}\Gamma \frac{1-{\xslash v}}{2} Q_v (
z ) e^{i (q + m_Q v + m_{Q^\prime} v^\prime ) \cdot z } |H_v >
\end{eqnarray}

For the meson state $H_v \equiv M_v = ( \bar{Q}_v q ) $ with $k =
\bar{\Lambda} v $ in HQEFT, the meson state $|H_v >$ can generally
be expressed as
\begin{eqnarray}
|H_v > & = & \int_k \int_{\tilde{k}} \delta ( \bar{\Lambda}_H v -
k - \tilde{k} ) b_v^+ ( k, s ) d_q^+ ( \tilde{k}, \tilde{s} )
\phi_{s \tilde{s}} ( k, \tilde{k}) | 0 > \nonumber \\
& \equiv &  \int_{\tilde{k}}  b_v^+ ( \bar{\Lambda}_H v -
\tilde{k} , s) d_q^+ ( \tilde{k}, \tilde{s} ) \phi_{s \tilde{s}}
(v\cdot \tilde{k} , \tilde{k}^2) | 0 > \nonumber \\
& \equiv & \int_{k}  b_v^+ ( k , s ) d_q^+ ( \bar{\Lambda}_H v -
k, \tilde{s} ) \phi_{s \tilde{s}} ( v\cdot k , k^2) | 0 >
\end{eqnarray}
here  we have used the definitions
\begin{eqnarray}
 & & \phi_{s \tilde{s}} (\bar{\Lambda}_H v - \tilde{k} , \tilde{k})
 \equiv \phi_{s \tilde{s}} (v\cdot \tilde{k} , \tilde{k}^2)
 \nonumber \\
 & & \phi_{s \tilde{s}} (k, \bar{\Lambda}_H v - k)
 \equiv  \phi_{s \tilde{s}} ( v\cdot k , k^2)
\end{eqnarray}
where $ \phi_{s \tilde{s} } ( k , \tilde{k} ) $ is the wave
function in momentum space. In this case, only the first term in
the transition matrix element $T_{\Gamma}^0$ becomes non-vanishing
\begin{eqnarray}
T_{\Gamma}^0 & & = <H' _{v'} | \int_z \bar{Q}_{v^\prime}^\prime (
z ) \frac{1+{\xslash v^\prime}}{2} \Gamma \frac{1+{\xslash v}}{2}
Q_v ( z )
e^{i (q - m_Q v + m_{Q^\prime } v^\prime  ) \cdot z } |H_v >\nonumber \\
& & = <0 | \int_{k^\prime} \int_{\tilde{k}^\prime} \phi_{s^\prime
\tilde{s}^\prime}^{\prime \dagger} ( k^\prime, \tilde{k}^\prime )
b_{v^\prime} ( k^\prime, s^\prime) d_q ( \tilde{k}^\prime,
\tilde{s}^\prime ) \delta ( \bar{\Lambda}_{H'} v^\prime - k^\prime
- \tilde{k}^\prime )
 \nonumber \\
& & \int_z \bar{Q}_{v^\prime}^\prime ( z ) \frac{1 + {\xslash v^\prime}}{2}
\Gamma \frac{ 1 + {\xslash v}}{2} Q_v ( z ) e^{ i ( q - m_Q v + m_{Q^\prime } v^\prime ) \cdot z}
\int_k \int_{\tilde{k}} b_v^+ ( k, s ) d_q^+ ( \tilde{k}, \tilde{s} )
\phi_{s \tilde{s}} ( k, \tilde{k}) | 0 > \nonumber \\
& & = <0 |\int_{k^\prime} \int_{\tilde{k}^\prime} \phi_{s^\prime
\tilde{s}^\prime}^{\prime \dagger} ( k^\prime, \tilde{k}^\prime )
d_q ( \tilde{k}^\prime, \tilde{s}^\prime )
\delta ( \bar{\Lambda}_{H'} v^\prime - k^\prime - \tilde{k}^\prime )\nonumber \\
& & ( - i {Z_2^\prime}^{-\frac{1}{2}} ) \int_y \bar{u} ( v^\prime, s^\prime )
 e^{ i k^\prime \cdot y} i {\xslash v^\prime} v^\prime \cdot \partial_y Q_{v^\prime}^\prime ( y )
 \int_z \bar{Q}_{v^\prime }^\prime ( z ) \frac{1 + {\xslash v^\prime }}{2}
 \Gamma \frac{1+ {\xslash v }}{2} Q_v ( z ) \nonumber \\
& & e^{ i ( q - m_Q v + m_{Q^\prime } v^\prime ) \cdot z }
\int_k \int_{\tilde{k}} ( - i Z_2^{-\frac{1}{2}} )
\int_x \bar{Q}_v ( x ) ( - i \overleftarrow{\partial}_x \cdot v {\xslash v} )
u ( v, s ) e^{- i k \cdot x }\nonumber \\
& & d_q^+ ( \tilde{k}, \tilde{s} ) \phi_{s \tilde{s} } ( k,
\tilde{k} ) \delta(\bar{\Lambda}_H v-k-\tilde{k})|0 >
\end{eqnarray}
where the definition of meson state eq.(5.16) and the LSZ
reduction formula for effective heavy quarks has been used in step
1 and step 2 respectively. Contracting the heavy quark fields
\begin{eqnarray}
& & Q_{v^\prime}^\prime ( y ) \bar{Q}_{v^\prime}^\prime ( z) \rightarrow
 \int_{l^\prime } e^{ - i l^\prime \cdot  ( y - z )}
 \frac{i}{{\xslash v^\prime} v^\prime \cdot l^\prime } \nonumber \\
& & Q_v ( z ) \bar{Q}_v ( x ) \rightarrow \int_l e^{- i l \cdot (
z - x ) } \frac{i}{{\xslash v} v \cdot l }
\end{eqnarray}
and integrating over $ x, y, z, l, l^\prime $ as well as using the
relation
\begin{eqnarray}
<0 | d_q ( \tilde{k}^\prime, \tilde{s}^\prime ) d_q^+ ( \tilde{k},
\tilde{s} ) | 0 > = \delta^3 ( \overrightarrow{\tilde{k}}^\prime -
\overrightarrow{\tilde{k}} ) \delta_{\tilde{s}^\prime \tilde{s} }
\frac{\tilde{k}^0}{m}
\end{eqnarray}
We obtain
\begin{eqnarray}
& & T_{\Gamma}^0   = < 0 | \int_{k^\prime} \int_{\tilde{k}^\prime
} \phi_{s^\prime \tilde{s}^\prime }^{\prime \dagger} ( k^\prime,
\tilde{k}^\prime ) \delta ( \bar{\Lambda}_{H'} v^\prime - k^\prime
- \tilde{k}^\prime )
( - i {Z_2^\prime }^{- \frac{1}{2}} ) \nonumber \\
& & \bar{u} ( v^\prime, s^\prime ) i \frac{1 + {\xslash v^\prime}}{2}
\Gamma i \frac{1+{\xslash v}}{2} u ( v, s ) ( -i Z_2^{- \frac{1}{2} } )
\int_k \int_{\tilde{k}} \delta ( q - m_Q v - k + m_{Q^\prime} v^\prime + k^\prime ) \nonumber \\
& & \delta ( \tilde{k}^\prime - \tilde{k} ) \phi_{s \tilde{s}} ( k, \tilde{k} )
\delta ( \bar{\Lambda}_H v -k - \tilde{k} ) | 0 > \nonumber \\
& & = \frac{\delta ( q - p + p^\prime )}{\sqrt{ Z_2^\prime Z_2 }}
\bar{u} ( v^\prime, s^\prime ) \frac{1+ {\xslash v^\prime }}{2}
\Gamma \frac{1+{\xslash v}}{2} u ( v, s )  \int_{\tilde{k}}
\phi_{s^\prime \tilde{s}}^{\prime \dagger} (v^\prime \cdot
\tilde{k}, \tilde{k}^2 ) \phi_{s \tilde{s} }(
v\cdot\tilde{k}, \tilde{k}^2 ) \nonumber \\
 & & \equiv  \frac{\delta ( q - p + p^\prime )}{\sqrt{ Z_2^\prime Z_2 }} \bar{u} ( v^\prime,
s^\prime ) \frac{1+ {\xslash v^\prime }}{2} \Gamma
\frac{1+{\xslash v}}{2} u ( v, s )  \int_{k} \phi_{s^\prime
\tilde{s}}^{\prime \dagger} (v^\prime \cdot k , k^2 ) \phi_{s
\tilde{s} }( v\cdot k, k^2 )
\end{eqnarray}

In general, the above wave functions can be written into the
following forms
\begin{eqnarray}
 & & \phi_{s \tilde{s} }( v\cdot\tilde{k}, \tilde{k}^2 ) =
   \bar{u}(v, s) i\gamma_5 v (\tilde{k}, \tilde{s} ) \varphi_P ( v\cdot\tilde{k}, \tilde{k}^2) \\
 & & \phi_{s^\prime \tilde{s}}^{\prime }(v^\prime \cdot \tilde{k}, \tilde{k}^2
 ) =  \bar{v}(\tilde{k}, \tilde{s} ) i \gamma_5 u(v', s')
 \varphi_P^{\prime }(v^\prime \cdot \tilde{k}, \tilde{k}^2)
\end{eqnarray}
for pseudoscalar mesons, and
\begin{eqnarray}
 & & \phi_{s \tilde{s} }(v\cdot\tilde{k}, \tilde{k}^2 ) =
 \bar{u}(v, s) {\xslash \epsilon}  v (\tilde{k}, \tilde{s} )
 \varphi_V ( v\cdot\tilde{k}, \tilde{k}^2)\\
 & & \phi_{s^\prime \tilde{s}}^{\prime }(v^\prime \cdot \tilde{k}, \tilde{k}^2
 )  =\bar{v}(\tilde{k}, \tilde{s} ) {\xslash \epsilon }u(v', s')
 \varphi_V^{\prime }(v^\prime \cdot \tilde{k}, \tilde{k}^2 )
\end{eqnarray}
for vector mesons. The spin-flavor symmetry implies that
$\varphi_V = \varphi_P = \varphi$.

 With the above formulation, the transition matrix can be
 simplified to be
 \begin{eqnarray}
T_{\Gamma}^0  & = & <H' _{v'} | \int_z \bar{Q}_{v^\prime}^\prime (
z ) \frac{1+{\xslash v^\prime}}{2} \Gamma \frac{1+{\xslash v}}{2}
Q_v ( z ) e^{i (q - m_Q v + m_{Q^\prime } v^\prime  ) \cdot z } |H_v >\nonumber \\
& = & \frac{\delta ( q - p + p^\prime )}{\sqrt{ Z_2^\prime Z_2 }}
Tr \left(  \bar{{\cal M}}_+(v') \Gamma {\cal M}_+(v)
\frac{1}{\bar{\Lambda}}\int_{\tilde{k}} ( 1 - \frac{
\tilde{{\xslash k}}}{m_q} ) \varphi^{\prime } (v^\prime
\cdot \tilde{k}, \tilde{k}^2 ) \varphi( v\cdot\tilde{k}, \tilde{k}^2 ) \right ) \nonumber \\
& = & \delta ( q - p + p^\prime ) \xi(v\cdot v')\ Tr \left(
\bar{{\cal M}}_+(v') \Gamma {\cal M}_+(v) \right)
 \end{eqnarray}
 where the ${\cal M}_+(v)$ is the spin wave functions
     \begin{eqnarray}
       {\cal M}_+(v)=\sqrt{\bar{\Lambda}}P_{+}
         \left\{
           \begin{array}{cl}
              -\gamma^{5}, & pseudoscalar \; meson \; \; P\\
              \epsilon\hspace{-0.15cm}\slash, & vector \; meson \; \; V
           \end{array}
         \right.
     \end{eqnarray}
  Here $\epsilon^{\mu}$ is the
  polarization vector of the vector meson. In obtaining the above results, we have used the
  following property
\begin{eqnarray}
  \frac{1}{\bar{\Lambda}}\int_{\tilde{k}} ( 1 - \frac{
\tilde{{\xslash k}}}{m_q} ) \varphi^{\prime } (v^\prime \cdot
\tilde{k}, \tilde{k}^2 ) \varphi( v\cdot\tilde{k}, \tilde{k}^2 ) =
\zeta (v\cdot v') ( 1 - \alpha {\xslash v} - \alpha' {\xslash v}'
)
 \end{eqnarray}
 and introduced the function $ \xi(v\cdot v') $ via the following
 definition
\begin{eqnarray}
  \xi(v\cdot v') = \frac{1}{\sqrt{ Z_2^\prime Z_2 }}\zeta (v\cdot v') ( 1 +  \alpha  + \alpha' )
\end{eqnarray}
which is the well-known universal Isgur-Wise function\cite{IW}.

 For the anti-meson state $H = \bar{M} = ( \bar{q} Q ) $, we have in
general
\begin{eqnarray}
| H_v > = \int_k \int_{\tilde{k}} \delta ( \bar{\Lambda}_H v - k -
\tilde{k} ) d_v^+ ( k, s ) b_q^+ ( \tilde{k}, \tilde{s} ) \phi_{s
\tilde{s}} ( k, \tilde{k}) | 0 >
\end{eqnarray}
where $ \phi_{s \tilde{s} } ( k, \tilde{k} ) $ is the wave
function in momentum space. In this case, only the second term in
the transition matrix element $T_{\Gamma}^0$ becomes
non-vanishing. Following the same procedure, one arrives at
 \begin{eqnarray}
T_{\Gamma}^0  & = & <H' _{v'} | \int_z \bar{Q}_{v^\prime}^\prime (
z ) \frac{1-{\xslash v^\prime}}{2} \Gamma \frac{1-{\xslash v}}{2}
Q_v ( z ) e^{i (q - m_Q v + m_{Q^\prime } v^\prime  ) \cdot z } |H_v >\nonumber \\
& = & \frac{\delta ( q - p + p^\prime )}{\sqrt{ Z_2^\prime Z_2 }}
Tr \left(  \bar{{\cal M}}_-(v') \Gamma {\cal M}_-(v)
\frac{1}{\bar{\Lambda}}\int_{\tilde{k}} ( 1 + \frac{
\tilde{{\xslash k}}}{m_q} ) \varphi^{\prime } (v^\prime
\cdot \tilde{k}, \tilde{k}^2 ) \varphi( v\cdot\tilde{k}, \tilde{k}^2 ) \right ) \nonumber \\
& = & \delta ( q - p + p^\prime ) \xi(v\cdot v')\ Tr \left(
\bar{{\cal M}}_-(v') \Gamma {\cal M}_-(v) \right)
 \end{eqnarray}
 where the ${\cal M}_-(v)$ is the spin wave function for
 anti-meson state
     \begin{eqnarray}
       {\cal M}_-(v)=\sqrt{\bar{\Lambda}}P_{-}
         \left\{
           \begin{array}{cl}
              -\gamma^{5}, & pseudoscalar \; meson \; \; P\\
              \epsilon\hspace{-0.15cm}\slash, & vector \; meson \; \; V
           \end{array}
         \right.
     \end{eqnarray}

 In general, we have
 \begin{eqnarray}
T_{\Gamma}^0  & = & <H' _{v'} |\bar{Q}_{v^\prime}^\prime \Gamma
Q_v  |H_v > \nonumber \\
& = &  \xi(v\cdot v')\ Tr \left( \bar{{\cal M}}'(v')\Gamma {\cal
M}(v) \right)
\end{eqnarray}
Where $\bar{\cal M}=\gamma^{0}{\cal M}^{\dagger}\gamma^{0}$ is the
spin-wave function in HQEFT. It is specified to ${\cal M} = {\cal
M}_+ $ for heavy mesons and ${\cal M}= {\cal M}_- $ for heavy
anti-mesons.  $\xi(\omega)$ is the Isgur-Wise function that
normalizes to unity at the point of zero recoil $\omega =1$, i.e.,
$\xi(1) = 1$.

  So far, we explicitly demonstrate the trace formula for the
  transition matrix elements in HQEFT at leading order.
  Generalization to higher orders is straightforward.

  \subsection{Transition Form Factors and $1/m_Q$ Expansions in HQEFT}

  We shall apply the trace formula obtained in the previous subsection
to evaluate the hadronic matrix elements. In general, the hadronic
matrix elements of vector and axial vector currents between
pseudoscalars and vector mesons are described by 18 transition
form factors
  \begin{eqnarray}
   &&\hspace{-0.7cm}<D(v^{\prime})\vert \bar{c}\gamma^{\mu} b \vert B(v)>
     =\sqrt{m_D m_B}[h_{+}(\omega)(v+v^{\prime})^{\mu}+h_{-}(\omega)(v-v^{\prime})^{\mu}] ,\nonumber\\
   &&\hspace{-0.7cm}<D^{\ast}(v^{\prime},\epsilon^{\prime})\vert \bar{c}\gamma^{\mu} b \vert B(v)>
         = i \sqrt{m_{D^{\ast}} m_B}
         h_{V}(\omega) \epsilon^{\mu \nu \alpha \beta} \epsilon^{\prime \ast}_{\nu}
         v^{\prime}_{\alpha} v_{\beta} ,\nonumber\\
   &&\hspace{-0.7cm}<D^{\ast}(v^{\prime},\epsilon^{\prime})\vert \bar{c}\gamma^{\mu} \gamma^{5}b \vert B(v)>
         = \sqrt{m_{D^{\ast}} m_B}
         [h_{A_1}(\omega)(1+\omega) \epsilon^{\prime \ast \mu}
       -h_{A_2}(\omega)(\epsilon^{\prime \ast} \cdot v)v^{\mu} \nonumber \\
  &&\hspace{1cm} -h_{A_3}(\omega)(\epsilon^{\prime \ast} \cdot v)v^{\prime\mu}], \nonumber\\
  &&\hspace{-0.7cm}<D^{\ast}(v^{\prime},\epsilon^{\prime})\vert \bar{c}\gamma^{\mu}b\vert
         B^{\ast}(v,\epsilon)>
       =\sqrt{m_{D^{\ast}} m_{B^{\ast}} }
    \{-(\epsilon\cdot \epsilon^{\prime\ast})[h_{1}(\omega)(v+v^{\prime})^{\mu}
       +h_{2}(\omega)(v-v^{\prime})^{\mu}] \nonumber\\
  &&\hspace{1cm} +h_{3}(\omega)(\epsilon^{\prime\ast}\cdot v)\epsilon^{\mu}
 +h_{4}(\omega)(\epsilon\cdot v^{\prime})\epsilon^{\prime\ast\mu}
       -(\epsilon\cdot v^{\prime})(\epsilon^{\prime\ast}\cdot v)
    [h_{5}(\omega)v^{\mu}+h_{6}(\omega)v^{\prime\mu}] \} ,\nonumber \\
&&\hspace{-0.7cm}<D^{\ast}(v^{\prime},\epsilon^{\prime})\vert
\bar{c}\gamma^{\mu} \gamma^{5}b\vert
        B^{\ast}(v,\epsilon)>
        =i\sqrt{m_{D^{\ast}} m_{B^{\ast}}}
        \{\epsilon^{\mu\nu\alpha\beta}\{\epsilon_{\alpha}\epsilon^{\prime\ast}_{\beta}
        [h_{7}(\omega)(v+v^{\prime})_{\nu}\nonumber\\
   &&\hspace{1cm}+h_{8}(\omega)(v-v^{\prime})_{\nu}]
   +v^{\prime}_{\alpha}v_{\beta}[h_{9}(\omega)(\epsilon^{\prime\ast}\cdot v)
       \epsilon_{\nu}+h_{10}(\omega)(\epsilon\cdot v^{\prime})
        \epsilon^{\prime\ast}_{\nu}] \} \nonumber\\
  &&\hspace{1cm}+\epsilon^{\alpha\beta\gamma\delta}\epsilon_{\alpha}
        \epsilon^{\prime\ast}_{\beta}v_{\gamma}v^{\prime}_{\delta}
       [h_{11}(\omega)v^{\mu}+h_{12}(\omega)v^{\prime\mu} ] \} .
 \end{eqnarray}

To relate the form factors $h_{i}(\omega)$ with the matrix
elements of operators in HQEFT, we reexpress the general form of
transition matrix elements in HQEFT via the $1/m_Q$ expansion
(\ref{matrixA}) to the following explicit forms
  \begin{eqnarray}
  \label{5.35}
     {\cal A} & = & <H^{\prime}_{v^{\prime}}\vert \QVBP \Gamma \QV \vert H_v>
         \nonumber \\
      & &  +\frac{1}{2\MQ}<H^{\prime}_{v^{\prime}}\vert \QVBP \Gamma
        \frac{1}{iv\cdot \partial}P_{+} [ D^2_{\bot}
        +\frac{i}{2}\sigma_{\alpha\beta}
        F^{\alpha\beta} ] \QV \vert H_v>   \nonumber\\
       &&
        -\frac{1}{2m_{Q'}}<H^{\prime}_{v^{\prime}}\vert \QVBP
        [\DCLS   +\frac{i}{2}\sigma_{\alpha\beta}
        F^{\alpha\beta} ] P^{\prime}_{+}
        \frac{1}{iv^{\prime}\cdot {\stackrel{\leftarrow}{\partial}}}\Gamma
         \QV \vert H_v>\nonumber\\
       &&
        -\frac{1}{4m^2_Q}<H^{\prime}_{v^{\prime}}\vert \QVBP \Gamma
        \frac{1}{iv\cdot \partial}P_{+}[iD^2_{\bot}(v\cdot D)
        -iD^{\alpha} v^{\beta}F_{\alpha\beta}+D^2_{\bot}\frac{1}{iv\cdot
        \partial}P_{+} D^2_{\bot} \nonumber\\
       &&
        -\frac{1}{2}\sigma_{\alpha\beta}F^{\alpha\beta}
        (v\cdot D) -\sigma^{\sigma\alpha}v^{\beta}D_{\sigma}F_{\alpha\beta}
        +D^2_{\bot}\frac{1}{iv\cdot \partial}P_{+}\frac{i}{2}
        \sigma_{\gamma\sigma}F^{\gamma\sigma}   \nonumber\\
       &&
        +\frac{i}{2}\sigma_{\alpha\beta}
        F^{\alpha\beta}\frac{1}{iv\cdot \partial}P_{+}D^2_{\bot}
        -\frac{1}{4}\sigma_{\alpha\beta}F^{\alpha\beta}
        \frac{1}{iv\cdot \partial } P_{+} \sigma_{\gamma\sigma}F^{\gamma\sigma}
        ]  \QV \vert H_v> \nonumber\\
      &&
        -\frac{1}{4\MQP^2}<H^{\prime}_{v^{\prime}}\vert \QVBP
        [i(v^{\prime}\cdot \DL)\DCLS
        -iF_{\alpha\beta}\DL^{\alpha}v^{\prime\beta}+\DCLS P^{\prime}_{+}
        \frac{1}{-iv^{\prime} \cdot {\stackrel{\leftarrow}{\partial}} }\DCLS  \nonumber\\
      &&
        -\frac{1}{2}(v^{\prime}\cdot \DL)\sigma_{\alpha\beta}F^{\alpha\beta}
        -F_{\alpha\beta}\DL_{\sigma}v^{\prime\beta}\sigma^{\sigma\alpha}
        +\frac{i}{2}\sigma_{\gamma\sigma}F^{\gamma\sigma}P^{\prime}_{+}
        \frac{1}{-iv^{\prime} \cdot {\stackrel{\leftarrow}{\partial}} }\DCLS \nonumber\\
      &&
        +\DCLS P^{\prime}_{+}\frac{1}{-iv^{\prime}\cdot {\stackrel{\leftarrow}{\partial}} }\  \frac{i}{2}
        \sigma_{\alpha\beta}F^{\alpha\beta}
        -\frac{1}{4}\sigma_{\gamma\sigma}F^{\gamma\sigma}P^{\prime}_{+}
        \frac{1}{-iv^{\prime}\cdot {\stackrel{\leftarrow}{\partial}} }\sigma_{\alpha\beta}F^{\alpha\beta}
        ]P^{\prime}_{+}\frac{1}{-iv^{\prime}\cdot
        {\stackrel{\leftarrow}{\partial}} }\Gamma  \QV \vert H_v>  \nonumber\\
      && +\frac{1}{4\MQP\MQ}<H^{\prime}_{v^{\prime}}\vert \QVBP [\DCLS
       P^{\prime}_{+}\frac{1}{-iv^{\prime}\cdot {\stackrel{\leftarrow}{\partial}} }\Gamma\frac{1}{iv\cdot \partial }
       P_{+}\DC^2   \nonumber\\
       &&+\DCLS P^{\prime}_{+}\frac{1}{-iv^{\prime}\cdot {\stackrel{\leftarrow}{\partial}} }
       \Gamma\frac{1}{iv\cdot \partial }
       P_{+}\frac{i}{2}\sigma_{\gamma\sigma}F^{\gamma\sigma}
       +\frac{i}{2}\sigma_{\alpha\beta}F^{\alpha\beta}P^{\prime}_{+}
       {\frac{1}{-iv^{\prime} \cdot {\stackrel{\leftarrow}{\partial}} }}
       \Gamma {\frac{1}{iv \cdot \partial }} P_{+}\DC^2  \nonumber\\
       &&
       -\frac{1}{4}\sigma_{\alpha\beta}F^{\alpha\beta}P^{\prime}_{+}
       {\frac{1}{-iv^{\prime} \cdot {\stackrel{\leftarrow}{\partial}} } }
       \Gamma\frac{1}{v\cdot \partial }P_{+}\sigma_{\gamma\sigma}
       F^{\gamma\sigma} ] \QV \vert H_v>
   \end{eqnarray}
with $\sigma^{\alpha
\beta}=\frac{i}{2}[\gamma^{\alpha},\gamma^{\beta}]$,
$P_{+}=\frac{1+v\hspace{-0.15cm}\slash}{2}$ and
$P^{\prime}_{+}=\frac{1+{v\hspace{-0.15cm}\slash}^{\prime}}{2}$
being project operators. Where $F^{\alpha
\beta}=[D^{\beta},D^{\alpha}]$ is the field strength of gluons
tensor.

One can now adopt the trace formulation approach to parameterize
the relevant matrix elements via Lorentz invariance
    \begin{eqnarray}
  & & <M^{\prime}_{v^{\prime}}\vert \QVBP\Gamma \QV \vert M_v> =-\xi(\omega)
     Tr[\bar{\cal M}^{\prime}\Gamma {\cal M}],
   \nonumber   \\
  & & <M^{\prime}_{v^{\prime}}\vert \QVBP\Gamma \frac{1}{iv\cdot \partial } P_{+} \DC^2
      \QV \vert M_v> = \kappa_1(\omega) \frac{1}{\bar{\Lambda}} Tr[\bar{\cal M}^{\prime}\Gamma {\cal M}],
\nonumber      \\
  & & <M^{\prime}_{v^{\prime}}\vert \QVBP\Gamma \frac{1}{iv\cdot \partial } P_{+} \frac{i}{2}
      \sigma_{\alpha\beta}F^{\alpha\beta}\QV \vert M_v> = - \frac{1}{\bar{\Lambda}}Tr[\kappa_{\alpha\beta}(v,v^{\prime})
      \bar{\cal M}^{\prime}\Gamma P_{+}\frac{i}{2}\sigma^{\alpha\beta}{\cal M}] , \nonumber\\
  & & <M^{\prime}_{v^{\prime}}\vert \QVBP\Gamma \frac{1}{iv\cdot \partial } P_{+}
     [i\DC^2(v\cdot D)-iD^{\alpha}v^{\beta} F_{\alpha\beta}]\QV \vert M_v>
      =-\varrho_1(\omega)\frac{1}{\bar{\Lambda}}Tr[\bar{\cal M}^{\prime}\Gamma {\cal M}] ,\nonumber \\
  & & <M^{\prime}_{v^{\prime}}\vert \QVBP\Gamma \frac{1}{iv\cdot \partial } P_{+}
      [\frac{1}{2}\sigma_{\alpha\beta}F^{\alpha\beta}(v\cdot D)+\sigma^{\sigma\alpha}
      v^{\beta}D_{\sigma} F_{\alpha\beta}]\QV \vert M_v> \nonumber\\
     && = - \frac{1}{\bar{\Lambda}}Tr[\varrho_{\alpha\beta}(v,v^{\prime})\bar{\cal M}^{\prime}\Gamma P_{+}
      \frac{i}{2}\sigma^{\alpha\beta}{\cal M}] ,\nonumber \\
  & & <M^{\prime}_{v^{\prime}}\vert \QVBP\Gamma \frac{1}{iv\cdot \partial } P_{+} \DC^2
     \frac{1}{iv\cdot \partial }P_{+}\DC^2  \QV \vert M_v>
      =-\chi_1(\omega)\frac{1}{\bar{\Lambda}^2} Tr[\bar{\cal M}^{\prime}\Gamma {\cal M}] , \nonumber\\
  & & <M^{\prime}_{v^{\prime}}\vert \QVBP\Gamma \frac{1}{iv\cdot \partial } P_{+}
      [\DC^2 \frac{1}{iv\cdot \partial }P_{+} \frac{i}{2} \sigma_{\alpha\beta}
      F^{\alpha\beta}
      + \frac{i}{2}\sigma_{\alpha\beta}F^{\alpha\beta} \frac{1}{iv\cdot \partial }
      P_{+} \DC^2 ] \QV \vert M_v>  \nonumber\\
        && =\frac{1}{\bar{\Lambda}^2} Tr[\chi_{\alpha\beta}(v,v^{\prime})\bar{\cal M}^{\prime}\Gamma P_{+}
      \frac{i}{2}\sigma^{\alpha\beta}{\cal M}],   \nonumber\\
  & & <M^{\prime}_{v^{\prime}}\vert \QVBP\Gamma \frac{1}{iv\cdot \partial } P_{+}
    \frac{i}{2}\sigma_{\alpha\beta} F^{\alpha\beta} \frac{1}{iv\cdot \partial } P_{+}
    \frac{i}{2}\sigma_{\gamma\sigma}F^{\gamma\sigma}\QV \vert M_v> \nonumber\\
  &&  =- \frac{1}{\bar{\Lambda}^2} Tr[\chi_{\alpha\beta\gamma\sigma}(v,v^{\prime})\bar{\cal M}^{\prime}\Gamma P_{+}
    \frac{i}{2}\sigma^{\alpha\beta} P_{+} \frac{i}{2}\sigma^{\gamma\sigma} {\cal M}] ,  \nonumber \\
  & & <M^{\prime}_{v^{\prime}}\vert \QVBP \DCLS P^{\prime}_{+} \frac{1}{-iv^{\prime}
  \cdot {\stackrel{\leftarrow}{\partial}}}
     \Gamma \frac{1}{iv\cdot \partial } P_{+} \DC^2 \QV \vert M_v>
     =-\eta_{1}(\omega) \frac{1}{\bar{\Lambda}^2} Tr[\bar{\cal M}^{\prime}\Gamma {\cal M}] ,\nonumber \\
  & & <M^{\prime}_{v^{\prime}}\vert \QVBP  \DCLS P^{\prime}_{+} \frac{1}{-iv^{\prime}
  \cdot {\stackrel{\leftarrow}{\partial}} }
     \Gamma \frac{1}{iv\cdot \partial } P_{+} \frac{i}{2} \sigma_{\alpha\beta}
     F^{\alpha\beta} \QV \vert M_v> \nonumber\\
    && =\frac{1}{\bar{\Lambda}^2} Tr[\eta_{\alpha\beta}(v,v^{\prime})
      \bar{\cal M}^{\prime}\Gamma P_{+}\frac{i}{2}\sigma^{\alpha\beta}{\cal M}], \nonumber \\
  & & <M^{\prime}_{v^{\prime}}\vert \QVBP \frac{i}{2} \sigma_{\alpha\beta} F^{\alpha\beta}
     P^{\prime}_{+} \frac{1}{-iv^{\prime} \cdot {\stackrel{\leftarrow}{\partial}} }
     \Gamma \frac{1}{iv\cdot \partial } P_{+} \frac{i}{2} \sigma_{\gamma\sigma}
     F^{\gamma\sigma} \QV \vert M_v>  \nonumber \\
    && =-\frac{1}{\bar{\Lambda}^2} Tr[\eta_{\alpha\beta\gamma\sigma}(v,v^{\prime})
      \bar{\cal M}^{\prime}\frac{i}{2}\sigma^{\alpha\beta} P^{\prime}_{+} \Gamma P_{+}
      \frac{i}{2}\sigma^{\gamma\sigma}{\cal M}]  .
  \end{eqnarray}
with $\omega=v\cdot v^{\prime}$. Where $\bar{\cal
M}=\gamma^{0}{\cal M}^{\dagger}\gamma^{0}$ is the spin-wave
function in HQEFT (it is specified to ${\cal M} = {\cal M}_+ $ for
heavy mesons and ${\cal M}= {\cal M}_- $ for heavy anti-mesons).
$\xi(\omega)$ is the Isgur-Wise function that normalizes to unity
at the point of zero recoil $\omega =1$, i.e., $\xi(1) = 1$.
$\kappa_{\alpha \beta}(v,v^{\prime})$, $\varrho_{\alpha
\beta}(v,v^{\prime})$, $\chi_{\alpha \beta}(v,v^{\prime})$,
$\eta_{\alpha \beta}(v,v^{\prime})$,
$\chi_{\alpha\beta\gamma\delta}(v,v^{\prime})$, and
$\eta_{\alpha\beta\gamma\delta}(v,v^{\prime})$ are the Lorentz
tensors. They can be decomposed into scalar form factors in terms
of the Lorentz tensor $g_{\mu\nu}$ as well as the Lorentz vectors
$\gamma_{\mu}$, $v_{\mu}$ and $v'_{\mu}$. Their Lorentz
decompositions are found to have the following general forms
     \begin{eqnarray}
       & &  \kappa_{\alpha \beta}(v, v^{\prime})=i\kappa_2(\omega) \sigma_{\alpha \beta}
         +\kappa_3(\omega) (v^{\prime}_{\alpha}\gamma_{\beta}-v^{\prime}_{\beta}
           \gamma_{\alpha}),\nonumber\\
  & & \varrho_{\alpha \beta}(v, v^{\prime})=i\varrho_{2}(\omega) \sigma_{\alpha \beta}
         + \varrho_3(\omega) (v^{\prime}_{\alpha}\gamma_{\beta}-v^{\prime}
           _{\beta} \gamma_{\alpha}),\nonumber\\
 & & \chi_{\alpha \beta}(v, v^{\prime})=i\chi_2(\omega) \sigma_{\alpha \beta}
         + \chi_3(\omega) (v^{\prime}_{\alpha}\gamma_{\beta}-v^{\prime}
           _{\beta} \gamma_{\alpha}), \nonumber\\
& & \eta_{\alpha \beta}(v, v^{\prime}) = i\eta_{2}(\omega)
\sigma_{\alpha \beta}
          +\eta_3(\omega) (v^{\prime}_{\alpha}\gamma_{\beta}-v^{\prime}
           _{\beta} \gamma_{\alpha}), \nonumber\\
& & \chi_{\alpha\beta\gamma\delta}(v, v^{\prime}) =
\chi_{4}(\omega)(g_{\alpha\gamma}
     g_{\beta\delta}-g_{\alpha\delta}g_{\beta\gamma})+\chi_{5}(\omega)
     \sigma_{\gamma\delta}\sigma_{\alpha\beta} \nonumber \\
     & & +i\chi_{6}(\omega)(g_{\alpha\gamma}
     \sigma_{\beta\delta} -g_{\beta\gamma}\sigma_{\alpha\delta} -g_{\alpha\delta} \sigma_{\beta\gamma}
     + g_{\beta\delta}\sigma_{\alpha\gamma})
     +\chi_{7}(\omega)(v^{\prime}_{\gamma} \gamma_{\delta}
     -v^{\prime}_{\delta} \gamma_{\gamma})(v^{\prime}_{\alpha}\gamma_{\beta}
     -v^{\prime}_{\beta}\gamma_{\alpha}) \nonumber \\
   & &  +\chi_{8}(\omega)(g_{\alpha\gamma} v^{\prime}_{\beta}v^{\prime}_{\delta}
    -g_{\beta\gamma} v^{\prime}_{\alpha}v^{\prime}_{\delta}
     -g_{\alpha\delta} v^{\prime}_{\beta}v^{\prime}_{\gamma}
     +g_{\beta\delta} v^{\prime}_{\alpha}v^{\prime}_{\gamma})
     \nonumber \\
   & &   +\chi_{9}(\omega)(g_{\alpha\gamma} v^{\prime}_{\beta}\gamma_{\delta}
     -g_{\beta\gamma} v^{\prime}_{\alpha}\gamma_{\delta}
     -g_{\alpha\delta} v^{\prime}_{\beta}\gamma_{\gamma}
   +g_{\beta\delta} v^{\prime}_{\alpha}\gamma_{\gamma}) \nonumber\\
    & &  +\chi_{10}(\omega)(g_{\alpha\gamma} \gamma_{\beta}v^{\prime}_{\delta}
     -g_{\beta\gamma} \gamma_{\alpha}v^{\prime}_{\delta}
     -g_{\alpha\delta} \gamma_{\beta}v^{\prime}_{\gamma}
     +g_{\beta\delta} \gamma_{\alpha}v^{\prime}_{\gamma}) \nonumber\\
   & &  +i\chi_{11}(\omega) \times (\sigma_{\alpha\gamma} v^{\prime}_{\beta}\gamma_{\delta}
     -\sigma_{\beta\gamma} v^{\prime}_{\alpha}\gamma_{\delta}
     -\sigma_{\alpha\delta} v^{\prime}_{\beta}\gamma_{\gamma}
     +\sigma_{\beta\delta} v^{\prime}_{\alpha}\gamma_{\gamma}) \nonumber\\
    & &  +i\chi_{12}(\omega)(\sigma_{\alpha\gamma} \gamma_{\beta}v^{\prime}_{\delta}
     -\sigma_{\beta\gamma} \gamma_{\alpha}v^{\prime}_{\delta}
   -\sigma_{\alpha\delta} \gamma_{\beta}v^{\prime}_{\gamma}
     +\sigma_{\beta\delta} \gamma_{\alpha}v^{\prime}_{\gamma}) , \nonumber \\
  & & \eta_{\alpha\beta\gamma\delta}(v, v^{\prime}) = \eta_{4}(\omega)(g_{\alpha\gamma}
     g_{\beta\delta}-g_{\alpha\delta}g_{\beta\gamma})+\eta_{5}(\omega)
     \sigma_{\gamma\delta}\sigma_{\alpha\beta} \nonumber\\
     & &     +i\eta_{6}(\omega)(g_{\alpha\gamma}
     \sigma_{\beta\delta}-g_{\beta\gamma}\sigma_{\alpha\delta} -g_{\alpha\delta} \sigma_{\beta\gamma}
+g_{\beta\delta}\sigma_{\alpha\gamma})
     +\eta_{7}(\omega)(v^{\prime}_{\gamma} \gamma_{\delta}
     -v^{\prime}_{\delta} \gamma_{\gamma})(v_{\alpha}\gamma_{\beta}
     -v_{\beta}\gamma_{\alpha}) \nonumber\\
    & &  +\eta_{8}(\omega)(g_{\alpha\gamma} v_{\beta}v^{\prime}_{\delta}
   -g_{\beta\gamma} v_{\alpha} v^{\prime}_{\delta}
     -g_{\alpha\delta} v_{\beta}v^{\prime}_{\gamma}
     -g_{\beta\delta} v_{\alpha}v^{\prime}_{\gamma}) \nonumber \\
    & &   +\eta_{9}(\omega)(g_{\alpha\gamma} v_{\beta}\gamma_{\delta}
     -g_{\beta\gamma} v_{\alpha}\gamma_{\delta}
     -g_{\alpha\delta} v_{\beta}\gamma_{\gamma}
 +g_{\beta\delta} v_{\alpha}\gamma_{\gamma}
     +g_{\alpha\gamma} \gamma_{\beta}v^{\prime}_{\delta}
     -g_{\beta\gamma} \gamma_{\alpha}v^{\prime}_{\delta}
     -g_{\alpha\delta} \gamma_{\beta}v^{\prime}_{\gamma} \nonumber\\
    & &  +g_{\beta\delta} \gamma_{\alpha}v^{\prime}_{\gamma})
   +i\eta_{10}(\omega)(v_{\beta} \gamma_{\delta}\sigma_{\alpha\gamma}
   -v_{\alpha} \gamma_{\delta}\sigma_{\beta\gamma}
     -v_{\beta} \gamma_{\gamma}\sigma_{\alpha\delta}
     +v_{\alpha} \gamma_{\gamma}\sigma_{\beta\delta}
     +\sigma_{\alpha\gamma}\gamma_{\beta}v^{\prime}_{\delta} \nonumber\\
     & & -\sigma_{\beta\gamma}\gamma_{\alpha}v^{\prime}_{\delta}
     -\sigma_{\alpha\delta}\gamma_{\beta}v^{\prime}_{\gamma}
     +\sigma_{\beta\delta}\gamma_{\alpha}v^{\prime}_{\gamma}).
    \end{eqnarray}
where the identity $ v_\alpha P_{+} \sigma^{\alpha \beta}{\cal
M}=0 $ has been used. Thus all matrix elements up to order of
$1/m_Q^2$ can be represented by the above scalar form factors.

In general, a heavy quark within a hadron cannot be on-mass shell
due to strong interactions among heavy quark and light quark as
well as soft gluons. The off-mass shellness of the heavy quark in
the heavy hadron is characterized by a residual momentum
$k=\bar{\Lambda}v + \tilde{k}$. The total momentum $p$ of the
heavy quark in a hadron may be written as:  $p=m_Q v + k =
\hat{m}_Q v + \tilde{k} $. Thus the residual momentum $k =
\bar{\Lambda}v + \tilde{k} $ of the heavy quark within a hadron is
assumed to comprise the main contributions of the light degrees of
freedom. Where $\tilde{k}$ is the part which depends on heavy
flavor and is suppressed by $1/m_Q$. With this picture the heavy
quark may be regarded as a `dressed heavy quark', and the heavy
hadron containing a single heavy quark is more reliable to be
considered as a dualized particle of a `dressed heavy quark'. For
this reason, the form factors defined in HQEFT should have a very
weak dependence on the light constituents of heavy hadrons. Thus
it is useful in HQEFT to define the `dressed heavy quark' mass as
\begin{equation}
\hat{m}_Q \equiv \lim_{m_{Q}\to \infty} m_{H} = m_Q
+\bar{\Lambda}.
\end{equation}
As the momentum $ \tilde{k} = p - \hat{m}_Q v$ carried by the
effective heavy quark field $Q_v$ within the heavy hadron is
expected to be much smaller than the binding energy
$\bar{\Lambda}$, so we can make the following expansion
  \begin{equation}
  \frac{1}{iv\cdot \partial }\rightarrow \frac{1}{v\cdot k} \to \frac{1}{\bar{\Lambda}+ v\cdot \tilde{k}}=
\frac{1}{\bar{\Lambda}}\left(1+O(\frac{v\cdot
\tilde{k}}{\bar{\Lambda}})\right)
  \sim \frac{1}{\bar{\Lambda}} .
  \end{equation}
Here $\bar{\Lambda}$ characterizes the effects of the light
degrees of freedom in the heavy hadron due to nonperturbative
effects.  Therefore, from the point of view in HQEFT, the `dressed
heavy quark'-hadron duality should become more reliable than the
naive heavy quark-hadron duality.

After completing the trace calculation, one can easily write down
the matrix elements of vector and axial vector currents between
pseudoscalar and vector mesons in terms of Lorentz scalar factors
$\kappa_i$ ,$\varrho_{i}$, $\chi_{i}$ and $\eta_{i}$. The most
general results at non-zero recoil are quite lengthy\cite{W1}. At
zero recoil point, one only needs to know four necessary form
factors which have the following simple forms
    \begin{eqnarray}
    h_{+}(1)&=&1+\frac{1}{8\bar{\Lambda}^2}(\frac{1}{m_b}-\frac{1}{m_c})^2
          [\ (\kappa_{1}+3\kappa_{2})^2
         -(\eta_{1}+3\eta_{2}-3\eta_{4}-9\eta_{5}-6\eta_{6})\  ] ,\nonumber\\
   h_{A_{1}}(1)&=&1+\frac{1}{8m_b^2\bar{\Lambda}^2} \{ [(\kappa_1+3\kappa_2)
         -\frac{m_b}{m_c}(\kappa_1  -\kappa_2)]^2
       -(\eta_{1}+3\eta_{2}-3\eta_{4}-9\eta_{5}-6\eta_{6}) \} \nonumber\\
  &&-\frac{1}{8m^2_c \bar{\Lambda}^2}
         (\eta_{1}-\eta_{2} -3\eta_{4} -\eta_{5}+2\eta_{6})
       +\frac{1}{4m_b m_c \bar{\Lambda}^2}
         (\eta_{1}+\eta_{2}+\eta_{4}+3\eta_{5}+2\eta_{6}) , \nonumber\\
     h_{1}(1)&=&1+\frac{1}{8\bar{\Lambda}^2}(\frac{1}{m_b}-\frac{1}{m_c})^2
       [ \ (\kappa_{1}-\kappa_{2})^2
       - (\eta_1-\eta_2-3\eta_{4}-\eta_{5}+2\eta_{6}) \  ], \nonumber \\
    h_{7}(1)&=&- 1 -\frac{1}{8\bar{\Lambda}^2}(\frac{1}{m_b} -\frac{1}{m_c})^2
      (\kappa_{1}-\kappa_{2})^2
       +\frac{1}{8\bar{\Lambda}^2}(\frac{1}{m^2_b}+\frac{1}{m^2_c})
       (\eta_1-2\eta_2-3\eta_{4}  \nonumber\\
&&     -\eta_{5}+2\eta_{6}) -\frac{1}{4m_b m_c
\bar{\Lambda}^2}(\eta_1-2\eta_2+\eta_{4}-\eta_{5}
     -2\eta_{6})
   \end{eqnarray}
where the normalization of $\xi(\omega)$ at zero recoil
$\xi(\omega=1)=1$ has been used. The meson masses are found from
the normalization condition to be
     \begin{eqnarray}
   m_{D(B)} & = &  m_{c(b)} +   \bar{\Lambda}_{D(B)}=
         \hat{m}_{c(b)} -(\frac{1}{m_{c(b)}}-\frac{\bar{\Lambda}}{2m^2_{c(b)}})
          (\kappa_1+3\kappa_2) \nonumber\\
         &  - & \frac{1}{2m^2_{c(b)} \bar{\Lambda}}(\varrho_{1} \bar{\Lambda}
          +3\varrho_{2} \bar{\Lambda} +\chi_1+3\chi_2 -3\chi_{4}
          -9\chi_{5}-6\chi_{6}) \nonumber \\
          & + & \frac{1}{4m^2_{c(b)} \bar{\Lambda}}(\eta_{1}+6\eta_{2}-3\eta_{4}
          -9\eta_{5}-6\eta_{6}) +O(\frac{1}{m_{c(b)}^3}) , \\
 m_{D^{\ast}(B^{\ast})} & = & m_{c(b)} +  \bar{\Lambda}_{D^{\ast}(B^{\ast})}=
          \hat{m}_{c(b)} -(\frac{1}{m_{c(b)}}-\frac{\bar{\Lambda}}{2m^2_{c(b)}})
          (\kappa_1-\kappa_2) \nonumber\\
          & - & \frac{1}{2m^2_{c(b)} \bar{\Lambda}}(\varrho_{1} \bar{\Lambda}
          -\varrho_{2} \bar{\Lambda}+\chi_1-\chi_2 -3\chi_{4}-\chi_{5}  + 2\chi_{6}) \nonumber\\
          & + & \frac{1}{4m^2_{c(b)} \bar{\Lambda}}(\eta_{1}-2\eta_{2}-3\eta_{4}
          -\eta_{5}+2\eta_{6}) +O(\frac{1}{m_{c(b)}^3})
     \end{eqnarray}

It is of interest to note that two form factors $h_{-}(\omega)$
and $h_{2}(\omega)$ in HQEFT vanish in the whole region of
momentum transfer, i.e., $h_{-}(\omega)=h_{2}(\omega)=0$. Such a
feature results from the fact that in HQEFT the operators in the
effective Lagrangian and effective current contain only terms with
even powers of ${\DSC}$ due to the contributions of
quark-antiquark interaction terms. Generally, the mesonic matrix
elements up to the second order power corrections can be described
by a set of 29 scalar form factors, which are universal functions
of the kinematic variable $\omega=v\cdot v^{\prime}$. Such a
number is less than the one introduced in the widely used heavy
quark effective theory containing no quark-antiquark coupling
terms, where 34 form factors are needed. All the interesting
features are attributed to the quark-antiquark interaction terms
in HQEFT. At zero recoil point, some of the form factors are
kinematically suppressed, only 15 universal form factors are
needed to describe the mesonic matrix elements up to order
$1/m^2_Q$. Where $\kappa_1$ and $\kappa_2$ characterize the
contributions of the order $1/m_Q$ operators at zero recoil. As
seen from the above results, the first order corrections to the
meson mass arise only from those two form factors. They play the
same roles as the parameters $\lambda_1$ and $\lambda_2$ defined
in the heavy quark effective theory without quark-antiquark
coupling terms. It is seen that the order $1/{\MQ}$ corrections in
all meson transition matrix elements of weak currents are
automatically absent at zero recoil in HQEFT after including the
contributions from the quark-antiquark interaction terms. The
absence of $1/m_Q$ order corrections at zero recoil was first
noticed in ref.\cite{Luke} for the matrix element in the $B\to
D^{\ast}$ transition, a detailed proof was made by adopting
equation of motion based on the widely used heavy quark effective
theory containing no quark-antiquark coupling terms. Nevertheless,
in that treatment unlike to HQEFT, some transition processes like
$B\to D$ transition remain receiving $1/m_Q$ corrections, which
actually displays an explicit difference between two effective
theories.

\subsection{Interesting Features in Applications of HQEFT}

It is seen from above analyzes that the theoretical framework of
HQEFT provides a powerful tool for systematically evaluating the
hadronic matrix elements via $1/m_Q$ expansion, it allows us to
explore its applications to heavy quark systems. Here we only
outline the most interesting features observed in the applications
of HQEFT.

It of interest to note that the quark-antiquark interacting terms
in HQEFT play an important role for understanding the heavy hadron
dynamics. Firstly, only the even powers of $\DSC$ appear in the
effective Hamiltonian for either quark fields or antiquark fields
when including the contributions from the quark-antiquark
interacting terms, which significantly simplifies the structure of
transition matrix elements in the $1/m_Q$ expansion of HQEFT. For
instance, the $1/m_Q$ corrections from current expansion and from
insertion of $1/m_Q$ order Lagrangian are attributed to the same
set of wave functions $\kappa_i(\omega)$ $(i=1,2,3)$, and the
$1/m_Q$ order corrections to meson masses are related to the zero
recoil values of wave functions $\kappa_1\equiv
\kappa_1(\omega=1)$ and $\kappa_2\equiv \kappa_2(\omega=1)$. Thus
at $1/m_Q$ order in HQEFT only 3 independent functions are
involved in both the weak transition matrix elements and ground
state meson masses. This feature allows us to determine certain
wave functions from the meson masses.

Secondly, it has been shown in HQEFT that the order $1/{\MQ}$
corrections in weak transition matrix elements between all ground
state mesons are automatically absent at zero recoil when
including the contributions from the quark-antiquark coupling
terms. Such a feature allows us to extract the CKM matrix element
$\vert V_{cb}\vert$ not only from the exclusive semileptonic decay
modes $B\rightarrow D^{\ast}l\nu$, but also from $B\rightarrow
Dl\nu$ decay modes. More precise extraction of $\vert V_{cb}\vert$
and $\vert V_{ub}\vert$ up to the $1/m_Q^2$ order corrections have
been carried out \cite{W1,W2,W3} within the framework of HQEFT.

Thirdly, as HQEFT can describe a slightly off-mass shell heavy
quark within a hadron, it is believed that HQEFT should also lead
to a consistent application for inclusive heavy hadron processes.
Introducing the concept of `dressed heavy quark' -hadron duality
based on HQEFT becomes more reasonable than a naive quark-hadron
duality. Here the ``dressed heavy quark" mass is defined as
$\hat{m}_Q=m_Q+\bar{\Lambda}$ which is related to the hadron mass
at a high order $1/m_Q$ corrections, explicitly, we have
$\hat{m}_Q = m_H [1 + O(1/m_Q^2) ]$. Thus the mass quantity
entering into the inclusive decay rates in the HQEFT formulation
is the well defined ``dressed heavy quark" mass
$\hat{m}_Q=m_Q+\bar{\Lambda}$ rather than the heavy quark mass
$m_Q$. As a consequence, the resulting inclusive decay rate
formulae of heavy hadrons are found to receive no $1/\hat{m}_Q$
order corrections when the decay rates are expressed in terms of
the physical heavy hadron masses. This is the most interesting
advantages of HQEFT, which allows us to perform the heavy quark
expansion at the point of the well-defined ``dressed heavy quark"
mass $\hat{m}_Q$ instead of the heavy quark mass $m_Q$. Such a
treatment successfully suppresses the next to leading order
contributions of the expansion and diminishes the possible large
uncertainties from the heavy quark mass $m_Q$. Of particular, we
are naturally led to a consistent explanation for the long term
puzzle of life time differences among bottom hadrons, i,e., the
resulting predictions for the ratios $\tau(B^0_s)/\tau(B^0)$ and
$\tau(\Lambda_b)/\tau(B^0)$\cite{W4} are remarkably consistent
with the experimental data. In fact, only in this sense, we can
simultaneously present a more precise and consistent determination
for $|V_{cb}|$ and $|V_{ub}|$ from the inclusive bottom hadron
decays. The numerical results for $|V_{cb}|$ and $|V_{ub}|$ were
presented in ref.\cite{W4} up to the order of $1/m_Q^2$
corrections as $1/m_Q$ order corrections are automatically absent
in HQEFT.

Finally, we would like to address that as $1/m_Q$ corrections can
systematically be computed and consistently be estimated by the
powers of $\bar{\Lambda}/m_Q$ in HQEFT. The leading behavior in
the heavy quark expansion of HQEFT can characterize the main
features for heavy quark systems. For instance, the scaling law
for the heavy meson decay constants was truly found in HQEFT to
hold in a good approximation for heavy bottom mesons\cite{W5}.

\section{Conclusions and Remarks}

We have shown that a large component QCD (LCQCD) with both large
component effective heavy quark and antiquark fields can directly
be derived from full QCD by integrating over the small components
of heavy quark and antiquark fields with $|{\bf p}| < E + m_Q$.
When the heavy quark is slightly off-mass shell with the residual
momentum $k = p - m_Q v $ satisfying, at the rest frame $v=
(1,0,0,0)$, the condition $|{\bf k}| \ll \sqrt{2 k^0 m_Q} \sim
\sqrt{2\bar{\Lambda}m_Q}$ for $k^0 \sim \bar{\Lambda}$, the
typical case is for the heavy-light hadron system with $ k^0 \sim
|{\bf k} | \sim \bar{\Lambda} \ll m_Q$, then LCQCD can well be
treated as a heavy quark effective field theory (HQEFT) via a
systematical heavy quark expansion in terms of $1/m_Q$ once the
contributions from the effective heavy quark-antiquark coupling
terms are considered. Its leading term characterizes the behavior
of heavy quarks in the infinite mass limit and explicitly displays
the heavy quark spin-flavor symmetry.

It has been seen that a basic theoretical framework of HQEFT can
be established via an alternative quantization procedure instead
of the usual canonical quantization, which includes the quantum
generators of Poincare group, the Hilbert and Fock space,
anticommutations and velocity superselection rule, propagator and
Feynman rules, finite mass corrections and renormalization in
HQEFT. The Lorentz invariance and discrete symmetries in HQEFT
have explicitly been checked. In addition to spin-flavor symmetry,
we have demonstrated some new symmetries in the infinite mass
limit. The trivialization of gluon couplings and decouple theorem
as well as the renormalization of Wilson loop have also been
discussed.

The weak transition matrix elements have been well defined in
HQEFT with a manifest spin-flavor symmetry in the infinite mass
limit. From the vector current conservation and the well defined
normalization of hadronic matrix element of vector current in the
full QCD theory and HQEFT, the heavy hadron mass can alternatively
be defined by the heavy quark mass and binding energy via $1/m_Q$
expansion, so that the heavy hadron masses are related to the
transition wave functions at zero recoil and can be used to
determined the numerical values of transition wave functions at
zero recoil. In particular, we have demonstrated that a simple
trace formulation for evaluating the transition matrix elements in
HQEFT is derivable by using the LSZ reduction formula. It has
explicitly been shown that the universal Isgur-Wise function of
transition matrix elements is related to the overlapping integral
between the wave functions of initial and final meson states. It
is of interest to note that the trace formulation approach is very
powerful for parameterizing the transition form factors via
$1/m_Q$ expansion in HQEFT.

In summary, a complete theoretical framework of HQEFT has been
established from QCD in the sense of effective quantum field
theory, where the effective heavy quark and antiquark fields have
been dealt with on the same footing in a fully symmetric way. The
large component effective heavy quark-antiquark coupling terms
have been shown to play an important role for analyzing $1/m_Q$
corrections when the ``longitudinal" and ``transverse" residual
momenta of heavy quark are at the same order of power counting in
the $1/m_Q$ expansion. It should be not surprised to understand
the results presented in \cite{W1,W2,W3,W4,W5,W6} for their
consistency with the experimental data and theoretical
expectations. We believe that more precise and consistent results
can be made from HQEFT which will further be tested by more
accurate experimental data based on B-factories and colliders.

{\bf Acknowledgments}:  This work was supported in part by the
National Science Foundation of China (NSFC) under the grant
10475105, 10491306, and the Project of Knowledge Innovation
Program (PKIP) of Chinese Academy of Sciences. The author wants to
thank Y.B. Zuo for checking the manuscript. This work was
partially based on author's early works as well as recent works
done with W.Y. Wang, Y.A. Yan and M. Zhong. He would also like to
thank C.H. Chang, K.T. Chao, Y.Q. Chen, Y.B. Dai, C.S. Huang, T.
Huang, C. Liu, C.D. Lu, C.F. Qiao, Z.J. Xiao, Y.D. Yang for
valuable discussions and conversations.

\end{document}